\documentclass[11pt]{article}
\usepackage{epsfig}
\usepackage{amsmath}
\usepackage[small,hang]{caption}
\usepackage{cite}
\usepackage{bm}
\newcommand{\CLrd}{{\rm d}}
\newcommand{\CLri}{\mathrm{i}}

\newcommand{\CLrS}{\mathrm{S}}
\newcommand{\CLrA}{\mathrm{A}}
\newcommand{\alphaS}{\alpha_\mathrm{s}}
\newcommand{\alphaem}{\alpha_\mathrm{em}}

\newcommand{\DfourR}{D_4^{\mathrm R}}
\newcommand{\DfourI}{D_4^{\mathrm I}}
\newcommand{\DelFR}{\Delta F^{\mathrm R}}
\newcommand{\DelFI}{\Delta F^{\mathrm I}}
\newcommand{\phifS}{\varphi_{4\mathrm S}}
\newcommand{\phifA}{\varphi_{4\mathrm A}}
\newcommand{\DelFtwo}{\Delta F^{(2)}}
\newcommand{\DelFthree}{\Delta F^{(3)}}
\newcommand{\DelFfour}{\Delta F^{(4)}}

\newcommand{\CLxP}{x_{_\mathrm{I\!P}}}
\newcommand{\CLket}[1]{\left|#1\right\rangle}

\newcommand{\CLbraket}[2]{\left\langle#1|#2\right\rangle}
\newcommand{\CLnumtextS}[1]{\ifcase#1 zero\or one\or two\or three\or four\or
  five\or six\or seven\or eight\or nine\or ten\or eleven\or twelve
  \else #1\fi}
\newcommand{\CLbfk}{{\bm k}}
\newcommand{\CLbfl}{{\bm l}}
\newcommand{\CLbfq}{{\bm q}}
\newcommand{\CLbfv}{{\bm v}}
\newcommand{\CLbfV}{{\bm V}}
\newcommand{\CLbfa}{{\bm a}}
\newcommand{\CLbfb}{{\bm b}}
\DeclareMathOperator{\CLtr}{tr}%
\DeclareMathOperator{\CLdisc}{disc}%
\numberwithin{equation}{section}
\begin{document}
\begin{titlepage}
\hfill
\hspace*{\fill}
\begin{minipage}[t]{4cm}
  DESY--99--068\\
  hep-ph/9906308
\end{minipage}
\vspace*{2.cm}
\begin{center}
  \begin{LARGE}
    {\bf An Estimate of Twist-Four\\
      Contributions at Small $x_\mathrm{B}$ and Low $Q^2$}\\
  \end{LARGE}
  \renewcommand{\thefootnote}{\fnsymbol{footnote}}
  \footnotetext[0]
  {Supported by the TMR Network ``QCD and Deep Structure of Elementary
    Particles''. One of us (C.B.) is supported by
    \textit{Graduiertenkolleg Theoretische Elementarteilchenphysik}.}
  \renewcommand{\thefootnote}{\arabic{footnote}}
  \setcounter{footnote}{0}
  \vspace{2.5cm}
  \begin{Large}
    {Jochen Bartels and Claas Bontus}\\
  \end{Large}
  \vspace{0.3cm}
  \textit{II.\ Institut f\"ur Theoretische Physik,
    Universit\"at Hamburg,\\ Luruper Chaussee 149,
    D-22761 Hamburg\footnote{email:
        bartels@x4u2.desy.de $\quad$ claas.bontus@desy.de}
    }
\end{center}
\vspace*{3.cm}
\begin{quotation}
  \noindent
  We present a first study of gluonic twist-four corrections to the deep
  inelastic structure function $F_2$ at small $x$ and small $Q^2$.  The
  calculations are based upon the double logarithmic approximation of the
  coupled twist-four evolution equations of the gluonic twist-four operators
  that are expected to be dominant at small $x$. We first review the analytical
  results which are presently available and define the framework of our
  calculation.  In the second part we discuss the connection with DIS
  diffractive dissociation which can be used to estimate the size of some of
  the twist-four corrections. In the final part we show, for three different
  choices of the input distributions, the relative magnitude of the
  leading-twist and the twist-four contributions.
\end{quotation}
\vfill
\end{titlepage}
%
%
\section{Introduction}
The observation that the rise of the deep inelastic structure function $F_2$ at
small $x$ starts already at surprisingly low values of $Q^2$, has stimulated
attempts to use the DGLAP evolution equations in the low $Q^2$ kinematical
regime \cite{GRV,MRS,BF}, where previously -- i.e.\ before the advent of HERA
-- perturbative QCD had not been expected to be applicable.  As a general
result of these studies, with a suitable choice of initial conditions (in
particular: not rising at small $x$) it is possible to describe, within the
framework of the leading-twist, next-to-leading-order QCD evolution equations,
the $Q^2$-evolution of the structure functions down to rather low $Q^2$ values.
However, recently \cite{Caldwell} evidence has been given that a rather sharp
transition away from this DGLAP evolution occurs between
$3\,$GeV$^2<Q^2<8\,$GeV$^2$, $10^{-4}<x<10^{-3}$.

The apparent success of DGLAP in the low $Q^2$ region, however, does not
provide much help for answering the question of how the transition from the
perturbative parton picture into the non-perturbative hadronic region at
$Q^2=0$ works. Most naively, one expects the breakdown of perturbative QCD to
be accompanied by the growth of non-leading perturbative terms. For deep
inelastic scattering, the expansion parameters are powers of the strong
coupling constant $\alphaS$ and powers of $m^2/Q^2$ (with $m$ denoting some
hadronic scale) which define the twist expansion. Whereas the former expansion
has already received some attention (e.g.\ the question of resummation in the
anomalous dimension for small $x$ \cite{EHW}), the r\^ole of non-leading twist
at small $x$ so far has not yet been addressed. In particular, if at low $Q^2$
the transition between the parton picture and Regge physics results from an
interplay of leading-twist and higher-twist corrections, one should expect to
see the presence of a negative twist-four term already {\it slightly above} the
transition. The results of \cite{Caldwell} indicate that twist-four might be
present even at not so small values of $Q^2$. If this is the case, existing
leading-twist parametrizations in the low $Q^2$-region need to be re-examined.
If, on the other hand, higher-twist really remains small down to unexpected low
values of $Q^2$, one should look for an explanation of this phenomenon. In any
case, a closer investigation of the r\^ole of higher-twist in the small $x$,
low $Q^2$-region is quite important.

Twist-four operators are known \cite{BFLK} to have their own evolution
equations. For the small $x$ domain, gluon operators are expected to be the
dominant ones. Whereas fermionic twist-four operators have been investigated in
some detail, very little is known about the gluonic ones.  Following the
arguments of \cite{EFP} we expect that twist-four gluonic contributions are
obtained by investigating the $Q^2$-dependence of QCD diagrams with two, three,
or four gluons in the $t$-channel.  Explicit calculations of these diagrams
have been done only for the small $x$ behavior of gluonic scattering
amplitudes: the BFKL \cite{BFKL} (ladder) amplitude, from which one obtains the
high energy behavior of the elastic scattering of a virtual photon on a gluon,
and diffractive deep inelastic scattering in the triple Regge region \cite{BW},
which allows to extract contributions with three and four gluons in the
$t$-channel. Expanding these amplitudes in inverse powers of $Q^2$ we expect to
reproduce the $Q^2$ evolution equations for twist-two and twist-four in the
double leading log approximation (DLA): leading logarithmic in both $\ln Q^2$
and $\ln 1/x$. A complete leading order calculation of the $Q^2$-evolution
equations of the twist-four gluon operators is missing, and it is very
important that this task will be addressed. In this paper we will adopt the
viewpoint that the combined limit: $x\to 0$, $Q^2\to \infty$ of diagrams with
two, three, or four gluons in the $t$-channel provides the DLA for the
twist-four gluon operators.

A numerical analysis of twist-four corrections to, say, $F_2$ follows the same
sequence of steps as the usual twist-two strategy: for an input scale $Q_0^2$
one has to choose an initial distribution (with a certain number of free
parameters), and by application of evolution equations one gets values of the
structure functions for larger $Q^2$. Since the number of twist-four operators
that mix under renormalization will, in general, be larger than in the
twist-two case, the number of free parameters in such an analysis will be
large, and a reliable combined analysis of twist-two and twist-four terms looks
extremely difficult. It might, therefore, be helpful to include information on
specific final states, which are known to belong to higher-twist. At present
the most promising candidate are diffractive final states, in particular the
diffractive production of vector mesons from longitudinal photons: this process
has been shown to be calculable within perturbative QCD \cite{R,BFGMS,CFS,MRT}.
It belongs to higher-twist, and the measured cross section is substantial
(about 20\,\% of the total diffractive cross section). Another potential source
of information are diffractive final states with hard jets.

In this paper we perform a first numerical estimate of gluonic twist-four
corrections at small $x$ and low $Q^2$. For this we make use of the currently
available elements of the twist-four evolution equations. In our scheme we,
first, review the existing small $x$ calculations of gluonic scattering
amplitudes, and extract from them the relevant pieces of the $Q^2$ evolution.
The currently existing calculations do not allow to go beyond DLA\@.  Generally
we are dealing with gluonic amplitudes which couple to the photon through a
fermion-loop.  As a first general result we find that, for a systematic study
of twist-four, we have to consider both the transverse and the longitudinal
polarization of the photon. The reason for this is the fact that in the
fermion-loop the twist-four contribution of the transverse photon looses one
power of $\ln Q^2$, whereas the longitudinal photon does not. This is quite
opposite to the leading-twist case where the transverse photon results in a
logarithm, and the longitudinal one does not.  Unfortunately, in DLA this
result forces us to compare contributions of different order in $\alphaS$: in
order to make a legitimate comparison between the twist-four transverse and
longitudinal contributions, we should know the longitudinal one up to the (in
the sense of DLA) next-to-leading terms. We, therefore, restrict ourselves to
ratios of twist-four corrections to DLA leading-twist results.  A second
general result of our investigation is the sign structure of the different
higher-twist terms. Both for the longitudinal and for the transverse twist-four
contributions we shall argue that there are four different pieces that have to
be taken into account, and they come with alternating signs. The total
twist-four contribution, therefore, depends crucially on the magnitude of each
individual term, i.e.\ the size of the initial condition.

The second purpose of this paper is the search for a practical procedure for
estimating the higher-twist contributions. We propose to use experimental data
of DIS diffractive final states. It is, therefore, necessary to first
investigate the r\^ole of higher-twist in diffractive dissociation and to
analyse the connection with the inclusive structure functions. In the second
part of this paper we, therefore, review diffractive dissociation, with
particular emphasis on higher-twist. For the determination of free parameters
in our numerical analysis we will make use only of longitudinal quark-antiquark
production; in order to understand the origin of the different signs, however,
we have to analyse also other diffractive final states. Contact with the
corrections to the structure functions is made through the AGK cutting rules
\cite{AGK}. We find that, in contrast to the naive expectation, we cannot
simply use the diffractive cross section to determine the higher-twist
corrections to $F_2$. There still remains an unknown piece in the initial
conditions which we presently cannot determine. Our central result on the AGK
rules is contained in eqn.\,\eqref{RelPhiDiff_PhiSA1}.

In the final part of this paper we present results of a numerical analysis.
We write
\begin{equation} \label{StrucFuncDefEq1}
  F_i(x,Q^2)=F_i^{\tau=2} + \Delta F_i \,, \qquad i=t,\ell
\end{equation}
and compare the different pieces. The twist-four correction $\Delta F_i$ (both
for $i=t$ and $i=\ell$) consists of four pieces (cf.\,\eqref{DeltaFsumDs1}),
and we will present numerical results for them, all normalized to
$F_t^{\tau=2}$. Because of the uncertainty in the initial conditions, we
consider three different choices, two with flat (in $x$) input distributions,
and one with a rising $x$-distribution.  As one of our main results for the
first two scenarios, we find, despite the strong cancellations among the
different twist-four contributions, that the total twist-four contribution is
not small at $Q^2=1\,$GeV$^2$ (between 15\% and 130\%). It is negative and
mainly due to the transverse photon. Unfortunately, the severe limitation due
to the DLA, together with the uncertainties of the initial conditions prevents
us from obtaining a more precise estimate.

The outline of this paper is as follows. In section~\ref{SecHigherTwist} we
give an overview of the higher-twist results which are presently available, and
of the formulae that we are using in our analysis. For a systematic study of
twist-four at small $x$ we find that it is necessary to consider contributions
of amplitudes with two, three and four gluons in the $t$-channel. We discuss
these contributions in several subsections and outline how we handle the
problem of running $\alphaS$. Next (section~\ref{SecDiffDiss}), we analyse the
r\^ole of higher-twist in DIS diffraction and we study the connection with the
deep inelastic structure function. Finally, in section~\ref{SecNumRes} we
describe the numerical calculations and present and discuss our results.
%
%
\section{Higher-Twist at small $x$} \label{SecHigherTwist}
We begin with a brief summary of the present theoretical status of the
evolution of higher-twist operators. A systematic classification of twist-four
operators can, in principle, be obtained using the arguments of R.\,K.\,Ellis
et.al.\ \cite{EFP}. However, in \cite{EFP} (and in several other studies of
higher-twist operators \cite{HT}) only fermionic twist-four operators have been
considered. At small $x$, on the other hand, it is the pure gluonic operators
that are expected to dominate, and a systematic discussion of twist-four gluon
operators is missing. General dimensional arguments lead to the following
operators. First, there are the four-gluon operators:
\begin{eqnarray} 
  \label{Four_G_Op1}
  {\rm Tr}\,\, F^{\mu_1 \alpha}\dots F^{\mu_i}_{\phantom{\mu_i}\alpha}
                  \dots F^{\mu_j \beta}\dots
               F^{\mu_n}_{\phantom{\mu_n}\beta}   \nonumber \\
  {\rm Tr}\,\, F^{\mu_1 \alpha} \dots F^{\mu_i \beta}
                  \dots F^{\mu_j}_{\phantom{\mu_j}\alpha}\dots
               F^{\mu_n}_{\phantom{\mu_n}\beta}   \\ 
  {\rm Tr}\,\, F^{\mu_1 \alpha} \dots F^{\mu_i \beta}
                  \dots F^{\mu_j}_{\phantom{\mu_j}\beta}\dots
               F^{\mu_n}_{\phantom{\mu_n}\alpha}  \nonumber
\end{eqnarray} 
(here $1<i,j<n$, and the dots denote products of covariant derivatives).  In
\cite{BFLK} these operators have been named ``quasipartonic'', and a general
discussion of their evolution has been given. In particular, it has been shown
that to leading order only two-body interaction kernels are needed. They are
the non-forward generalizations of the AP splitting functions. An analysis of
the small $x$ evolution equations in DLA can be found in \cite{B}. In addition
to the quasipartonic operators \eqref{Four_G_Op1}, there exists another gluonic
twist-four operator which is contained in the BFKL equation:
\begin{equation}
  \label{Two_G_Op1}
  {\rm Tr}\,\, F^{\mu_1 \alpha} D^{\mu_2}\dots D^{\mu_{n-1}}
    F^{\mu_n}_{\phantom{\mu_n}\alpha}
    g_{\mu_i\, \mu_j}^{\perp}\;+\; {\rm Perm} \,\, .
\end{equation}
In DLA the anomalous dimension of this operator can be derived from the BFKL
equation; the recently derived \cite{FL} second order corrections to the BFKL
kernel allow to go beyond DLA\@. In general, the twist-four operators
\eqref{Four_G_Op1} and \eqref{Two_G_Op1} should mix, i.e.\ there should be
off-diagonal elements in the anomalous dimension matrix. These off-diagonal
elements are known in DLA-accuracy \cite{B}.

In the discussion below we shall see that the operators \eqref{Four_G_Op1} are
the most interesting ones at small $x$: they result in contributions which rise
stronger at small $x$ than the contributions of the leading-twist gluon
operator. Therefore, at sufficiently small $x$, this operator will catch up
with the leading-twist contributions, despite its $1/Q^2$ suppression. On the
other hand, power counting in $\alphaS$ will show that \eqref{Four_G_Op1} is
suppressed compared to \eqref{Two_G_Op1}: again, because of the strong small
$x$ rise in \eqref{Four_G_Op1}, \eqref{Four_G_Op1} will eventually dominate
over \eqref{Two_G_Op1}. The $\alphaS$-suppression of\,\eqref{Four_G_Op1} will,
however, force us to consider higher order corrections to \eqref{Two_G_Op1}.

In the following we shall list the presently available results for the
operators \eqref{Four_G_Op1} and \eqref{Two_G_Op1}. They do not allow us to go
beyond the DLA approximation. It remains a major theoretical effort that has to
be made in order to calculate the full leading order evolution kernels of the
gluonic operators.
%
%
\subsection{The Two-Gluon Operator}
\begin{figure}
  \begin{center}
    \input{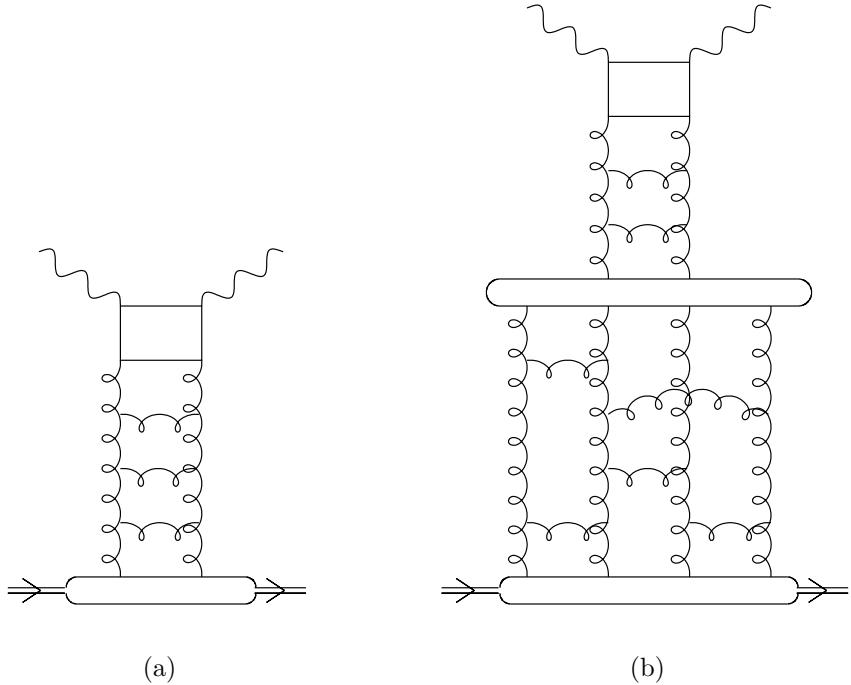}%
    \caption{(a) The two-gluon amplitude and (b) $\DfourI$, the part of
        the four-gluon amplitude which is \textit{irreducible} with respect of
        reggeization.}\label{G_Lad1}
  \end{center}
\end{figure}

As we have said before, the DLA approximation of the anomalous dimension of the
twist-four two-gluon operator can be obtained from the BFKL equation. Let 
$D_2^i(x,Q_0^2/Q^2)$ ($i=t,\ell$) denote the imaginary part of the scattering
amplitude of the photon (virtuality $Q^2$) off the gluon (virtuality $Q_0^2$).
It is convenient to write this amplitude as a double Mellin-transformation 
\begin{equation} \label{DoubleMelTrans1}
  D_{2}^i(x,Q_0^2/Q^2)= \int\! \frac{\CLrd\omega}{2\pi\CLri} \,\,
  x^{-\omega} \!\!\!\!\!\!\int\limits_{-1/2-\CLri\infty}^{-1/2+\CLri\infty}
  \frac{\CLrd\nu}{2\pi\CLri} \,
  \left(\frac{Q_0^2}{Q^2}\right)^{-\nu} D_{2}^i(\omega,\nu) \,.
\end{equation}
In the BFKL approximation, the amplitude $D_{2}^i(\omega,\nu)$ can be written
as
\begin{equation}
  D_{2}^i(\omega,\nu)=D_{2,0}^i(\nu) \frac{1}{\omega - \chi(\nu)} \,,
\end{equation}
where $D_{2,0}^i$ denotes the quark-loop at the upper end of the 
gluon ladder (fig.\,\ref{G_Lad1}a), and the BFKL characteristic function has
the form
\begin{equation}
  \chi(\nu)=\frac{N_\mathrm c\alphaS}\pi
  \left[ 2\psi(1)-\psi(\nu+1)
    -\psi(-\nu)  \right] \, .
\end{equation}
It has poles at integer values of $\nu$, and the anomalous dimensions of the
leading-twist two-gluon operator and the twist-four two-gluon operator are
obtained by solving the equation
\begin{equation}
  0=\omega - \chi(\nu)
\end{equation}
near the poles at $\nu=-1$ and $\nu=-2$, resp.:
\begin{eqnarray}
  \gamma^{\tau=2}(\omega) &=&\frac{N_\mathrm c}{\pi\omega}\alphaS 
    + \mathcal O\left( \left(\frac{\alphaS}{\omega}\right)^4 \right) \\
  \label{BFKLanomDims1}
  \gamma^{\tau=4}(\omega) &=& \frac{N_\mathrm c}{\pi\omega}\alphaS -
    2 \left( \frac{N_\mathrm c\alphaS}{\pi\omega} \right)^2 
    + \mathcal O\left( \left(\frac{\alphaS}{\omega}\right)^3 \right) \,.
\end{eqnarray}
Using the recently published NLO calculations \cite{FL} it is possible to
derive the singular part of the NLO corrections to the anomalous dimensions,
i.e.\ corrections of the form $\alphaS^2 \frac{\mathrm{const}}{\omega}$.

The coefficient functions are derived from the quark-loop
$D_{2,0}^i(\omega,\nu)$.  We begin with the result for the transverse photon
\cite{BW,WuestPHD}:
\begin{equation} \label{D20nu2}
  \frac1\omega D_{2,0}^t(\omega,\nu)=
  \frac1\omega \sum_f e_f^2\frac{\sqrt8}{2\pi}\alphaS
  \frac{\pi^2}{4}\frac{(\nu-1)(\nu+2)}{(\nu-\frac12)(\nu+\frac12)(\nu+\frac32)}
  \frac{\sin\left[\pi(\nu+\frac12)\right]}{\cos^2\left[\pi(\nu+\frac12)\right]}
  \, .
\end{equation}
Similarly, for the longitudinal photon one has:
\begin{equation} \label{D20nu1}
  \frac1\omega D_{2,0}^\ell(\omega,\nu)=
  \frac1\omega \sum_f e_f^2\frac{\sqrt8}{2\pi}\alphaS
  \frac{\pi^2}{4} \frac{\nu(\nu+1)}{(\nu-\frac12)(\nu+\frac12)(\nu+\frac32)}
  \frac{\sin\left[\pi(\nu+\frac12)\right]}{\cos^2\left[\pi(\nu+\frac12)\right]}
  \, .
\end{equation}
By expanding eqns.\,\eqref{D20nu2} and \eqref{D20nu1} in Laurent-series,
\begin{equation}
  D_{2,0}^{t,\ell}(\omega,\nu)=
  \sum_f e_f^2\frac{\sqrt8}{2\pi}\alphaS
  \sum_{n=-\infty}^{\infty} \left(
    \frac{b_n^{t,\ell}}{(\nu+n)^2} +\frac{a_n^{t,\ell,}}{\nu+n}
  \right) \, ,
\end{equation}
and considering the residues at negative integer values of $\nu$ we obtain the
twist expansion in powers of $Q_0^2/Q^2$.  The coefficients $a_n$ and $b_n$ for
leading- and next-to-leading twist are listed in table~\ref{D20expcoeff1}.
Since the double poles in the $b_n$-terms correspond to logarithms of $Q^2$
while the single poles in the $a_n$-terms correspond to constants, we recognize
with the help of table~\ref{D20expcoeff1} that, for leading-twist, the
longitudinal contribution is sub-leading at large $Q^2$ (compared to the
transverse cross section). For twist-four we have the opposite situation: the
longitudinal contribution has the $\ln Q^2/Q_0^2$ , whereas the transverse
cross section is suppressed. Moreover, we note the sign structure of the
coefficients in table~\ref{D20expcoeff1}: the twist-four contribution of the
longitudinal cross section comes with a negative sign.

\begin{table}
  \begin{center}
    \begin{tabular}{c|cc|cc}
      $n$ & $a_n^\ell$ & $b_n^\ell$ & $a_n^t$ & $b_n^t$ \\[2pt]
      \hline
      & & & & \\[-7pt]
      1 & $\frac23$ & $0$ & $\frac{14}{9}$ & $\frac43$ \\[5pt]
      2 & $-\frac{94}{225}$ & $-\frac{4}{15}$ & $\frac25$ & $0$ \\
    \end{tabular}
    \caption{Twist-expansion coefficients of $D_{2,0}$ for leading- and
      next-to-leading twist.}
    \label{D20expcoeff1}
  \end{center}
\end{table}
For our twist-four analysis we perform the computations in DLA and, therefore,
consider only amplitudes proportional to $b_2^\ell$ and $a_2^t$ for the
longitudinal and transverse contributions, respectively. Taking into account
the evolution in $Q^2$, we get, for each rung, a factor
$\gamma_2/\omega(\nu+2)$, where\footnote{Within our conventions the singular
  part of the gluon anomalous dimension is given by $\gamma_2/\omega$ and not
  by $\gamma_2$ itself.}
\begin{equation}
  \gamma_2\equiv\frac{N_\mathrm c\alphaS}{\pi} \,.
\end{equation}
Performing the $\nu$-integration we end up with
\begin{equation} \label{D2trans1}
  D_2^t(\omega,Q_0^2/Q^2)^{\tau=4}=
  a_2^t \sum_f e_f^2 \frac{\sqrt8}{2\pi} \alphaS
  \left( \frac{Q_0^2}{Q^2}\right)^2 \frac1{\omega}
  \exp\left(\frac{\gamma_2}{\omega} \ln(Q^2/Q_0^2)\right)
  , \qquad a_2^t = \frac25
\end{equation}
for the transverse amplitude, and 
\begin{equation} \label{D2long1}
  D_2^\ell(\omega,Q_0^2/Q^2)^{\tau=4}=
  b_2^\ell \sum_f e_f^2 \frac{\sqrt8}{2\pi} \frac{\alphaS}{\gamma_2}
  \left(\frac{Q_0^2}{Q^2}\right)^2 \left[\exp\left(\frac{\gamma_2}{\omega}
      \ln(Q^2/Q_0^2) \right)-1\right],
   \qquad b_2^\ell = -\frac4{15}
\end{equation}
for the longitudinal one. As stated before, the longitudinal amplitude has an
extra factor $\ln Q^2/Q_0^2$, in comparison with the transverse case: in the
DLA approximation, therefore, we have to be careful when comparing the
transverse amplitude with the longitudinal one.  However, the extra logarithm
only emphasises the dominance of the longitudinal amplitude at large $Q^2$. In
the low $Q^2$ region, there is no reason why the (twist-four) part of the
transverse cross section should be smaller than the longitudinal one. We will,
therefore, be slightly inconsistent with the counting of `leading' and
`sub-leading'.

Finally, we note the connection between the amplitudes $D_2^{t,\ell}$ and the
twist-four corrections to the structure functions:
\begin{eqnarray} \label{D2final}
  \Delta F_{t, \ell}^{(2)}(x,Q^2)&= &\frac{1}{2\sqrt{8}} \frac{Q^2}{Q_0^2} \int
  \frac{\CLrd\omega}{2\pi\CLri} \,\,
  x^{-\omega} \!\!\!\!\!\!\int\limits_{-1/2-\CLri\infty}^{-1/2+\CLri\infty}
  \frac{\CLrd\nu}{2\pi\CLri} \,
  \left(\frac{Q_0^2}{Q^2}\right)^{-\nu} 
  D_2^{t,\ell} (\omega, \nu) 
  \varphi_2(\omega,\nu)^{\tau=4} \,,
\end{eqnarray}
where the superscript `2' on the l.h.s.\ refers to the `2-gluon' ladder.  In
the subsequent sections we will add further terms to $\Delta F_i$. Within our
representations the twist-four part is obtained by evaluating the residue at
$\nu=-2$ (from $\nu=-1$ we get the leading-twist term).
$\varphi_2(\omega,\nu)$ denotes the initial condition: in the context of the
BFKL equation (which is an evolution equation in $y=\ln 1/x$) $\varphi_2$ would
define, at fixed $x_0$, the distribution in $Q^2$. However, in our DLA
approximation, where the BFKL equation coincides with the small $x$ limit of
the DGLAP evolution equation, $\varphi_2$ denotes, at fixed scale $Q_0^2$, the
initial distribution in $x$. In the ($\omega$, $\nu$)-representation,
$\varphi_2$ depends upon both $\nu$ and $\omega$ and contains both
leading-twist and higher-twist: the $\omega$ singularity structure near
$\nu=-1$ defines the $x$-shape of the leading-twist initial distribution of the
gluon structure function, and the behavior near $\nu=-2$ determines the input
to the twist-four correction.  As usual, both input distributions have to be
modelled and cannot be computed within pQCD\@. In the following, unless stated
differently, we will use $\varphi_2$ to denote the leading-twist initial
distribution.

Our discussion below will show that the situation with the two-gluon amplitude,
in fact, is more complicated. In particular, we shall argue that the
reggeization of the gluon provides extra contributions to the two-gluon
operator. Namely, in the next subsection we will turn to a discussion of the
four-gluon operators \eqref{Four_G_Op1}. To lowest order in $\alphaS$, it
involves diagrams in which four gluons are exchanged between the quark-loop and
the target proton (fig.\,\ref{FGLeadOrd1}). These diagrams are suppressed by
one factor $\alphaS$ compared to the two-gluon operator \eqref{Two_G_Op1}. At
small $x$, however, this suppression is compensated by the stronger rise of the
four-gluon operator contributions. When higher order gluon diagrams are
included, it will turn out that, at small $x$, the sum of all these diagrams
has to be split up into two separate classes. The first one corresponds to the
expected four-gluon operator (which, however, mixes with the two-gluon operator
\eqref{Two_G_Op1}). For this piece, the $\alphaS$-suppression (relative to the
twist-four two-gluon operator) is compensated by the stronger rise at small
$x$. However, in addition to this contribution to the four-gluon operator there
is a second set which results from the reggeization of the gluon. This
contribution has the same form of gluon ladders as the BFKL equation, except
that at the lower end the reggeizing gluon lines split up into two or more
gluon propagators (fig.\,\ref{D4Rfig1}). Starting from the lower end, the
four-gluon state immediately turns into a state of two reggeized gluons. It,
therefore, has the same $Q^2$-dependence as the BFKL amplitude, but counting
powers of $\alphaS$, it is of higher order than the BFKL ladder discussed
above. The lowest order diagram is the four-gluon diagram in
fig.\,\ref{FGLeadOrd1}, but in higher order it turns into a (higher order)
contribution to the initial distribution of the two-gluon operator. We will
show that this contribution can be related to the measured diffractive cross
section. After the discussion in the next subsection we, therefore, will, once
more, return to the two-gluon operator \eqref{Two_G_Op1} and add two more
separate contributions.

To summarize the main results of this subsection, starting from the BFKL
amplitude we have collected the known results on the twist-four two-gluon
operator \eqref{Two_G_Op1}. Rather than writing the BFKL amplitude as an
evolution equation we have used the closed expression in the
$(\omega,\nu)$-representation. Presently, only expressions within the DLA are
completely available. In order to go beyond the DLA it will be necessary to
perform a more complete calculation. For the anomalous dimension in
\eqref{BFKLanomDims1}, the next-to-leading order can be extracted from the
recent results of the second order BFKL kernel \cite{FL}. For the coefficient
functions (\eqref{D20nu2}, \eqref{D20nu1}) such a calculation has not yet been
performed.
%
%
%
%
\subsection{The Four-Gluon Operator}
\begin{figure}
  \begin{center}
    \input{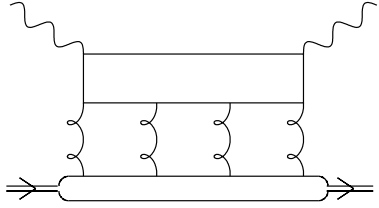}
    \caption{The four-gluon amplitude at leading order. The fermion-loop
      symbolizes the sum over all Feynman-diagrams with all possible couplings
      of the four $t$-channel gluons.} \label{FGLeadOrd1}
  \end{center}
\end{figure}

While the contributions of the two-gluon twist-four operator \eqref{Two_G_Op1}
could be derived with the help of the BFKL equation, the four-gluon operators
\eqref{Four_G_Op1} require a separate study. The simplest diagrams are shown in
fig.\,\ref{FGLeadOrd1}. In the limit of small $x$, these diagrams (together
with higher order corrections) have been investigated in \cite{B,BW}, and we
simply summarize the results and relate them to the operators
\eqref{Four_G_Op1} and \eqref{Two_G_Op1}.

\begin{figure}
  \begin{center}
    \input{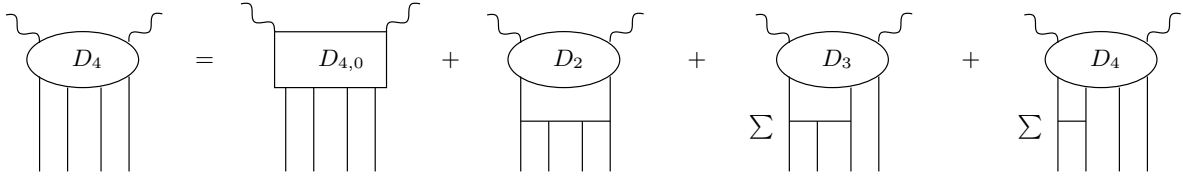}
    \caption{The integral equation for $D_4$. The sums denote couplings to the
      gluon lines in all possible ways.}\label{FacDiags1}
  \end{center}
\end{figure}

In the investigation of \cite{B,BW}, the starting point was the triple Regge
limit of a six-point amplitude; from this an amplitude $D_4$ was defined which
describes the coupling of four $t$-channel gluons to two virtual photons. We
recall only those results which are relevant for our analysis. The four-gluon
amplitude $D_4$ satisfies an integral equation which we illustrate with the
help of fig.\,\ref{FacDiags1}. As one main result of ref.\cite{BW}, it was
shown that $D_4$ should be written as a sum of two pieces, the
\textit{reggeizing} terms $\DfourR$, which consist of terms proportional $D_2$
(eqn.\,\eqref{D4R} below), and terms which are \textit{irreducible} with
respect of reggeization $\DfourI$:
\begin{equation} \label{D4sumD4R_D4I}
  D_4 = \DfourR + \DfourI \, .
\end{equation}
The two terms on the r.h.s.\ can also be distinguished by their symmetry
properties: whereas $\DfourI$ is totally symmetric under permutation of the
outgoing gluons, the term $\DfourR$ has mixed symmetry properties.  In the
following we shall discuss these contributions in some detail.  In particular,
we shall interpret them in terms of the operators \eqref{Four_G_Op1} and
\eqref{Two_G_Op1}. In order to avoid a too complicated notation, we will from
now on suppress the reference to the photon polarization.

We choose our conventions such that the connection with the cross section
$\sigma_{\gamma^*p}$ is as follows:
\begin{equation}  \label{D4crosssection}
  \sigma_{\gamma^*p} = -\frac{4\pi^2 \alphaem}{Q^2} \frac{1}{128\pi^2}
  \int\!\frac{\CLrd\omega}{2\pi\CLri} \left(\frac{1}{x} \right)^{\omega}
  \int\!\frac{\CLrd\nu}{2\pi\CLri} \left(\frac{Q_0^2}{Q^2} \right)^{-\nu-1}
  D_4^{abcd}\, \varphi_4^{abcd} \,.
\end{equation}
In the following we will describe this formula in more detail. In particular,
the overall minus sign in front will be derived with the help of the AGK rules
\cite{AGK}.

We begin with the \textit{irreducible} part. Since $\DfourR$ has a particularly
simple form (fig.\,\ref{D4Rfig1} and eqn.\,\eqref{D4R} below), we simply
subtract it on both sides of the equation in fig.\,\ref{FacDiags1} and make use
of the integral equation for $D_2$. Regrouping the terms on the r.h.s.\ of
fig.\,\ref{FacDiags1}, one arrives at the following integral equation:
\begin{equation}
  \label{Bet_Salp1}
  \omega D_4^{\mathrm I;abcd}(\CLbfk_1,\CLbfk_2,\CLbfk_3,\CLbfk_4) = 
  \left[D_2\otimes V^{abcd} + D_4^{\mathrm I;abcd} \otimes\sum 
    (K_{2\to2}+\alpha-1) \right](\CLbfk_1,\CLbfk_2,\CLbfk_3,\CLbfk_4) \, .
\end{equation}
Here $\bm{k}_1,\dots,\bm{k}_4$ are the transverse components (in the Sudakov
decomposition) of the momenta of the four gluons at the lower end in
fig.\,\ref{G_Lad1}b, and $a,b,c,d$ are the corresponding color labels.
$K_{2\to 2}$ and $\alpha$ are the common BFKL kernel and the gluon trajectory
function, and the sum has to be taken over all possible pairs of gluons.  $V$
represents the transition vertex from $2\to 4$ gluons (fig.\,\ref{G_Lad1}b),
\begin{alignat}{2} \label{Vdef1}
  V^{abcd}=\frac{g^2}{12\sqrt2}\Bigl\{ & 
  \phantom{+}\,\delta^{ab}\delta^{cd}\, && \bigl[
    G(\CLbfk_1,\CLbfk_3)+G(\CLbfk_2,\CLbfk_3)+G(\CLbfk_1,\CLbfk_4)+
    G(\CLbfk_2,\CLbfk_4) - \notag\\[-6pt] &&&
    -G(\CLbfk_1,\CLbfk_3+\CLbfk_4)-G(\CLbfk_2,\CLbfk_3+\CLbfk_4)-
    G(\CLbfk_3,\CLbfk_1+\CLbfk_2) - \notag\\ &&&
    -G(\CLbfk_4,\CLbfk_1+\CLbfk_2)+G(\CLbfk_1+\CLbfk_2,\CLbfk_3+\CLbfk_4)
    \bigr] + \notag \\ &
  +\delta^{ac}\delta^{bd} && \bigl[
    G(\CLbfk_1,\CLbfk_2)+G(\CLbfk_2,\CLbfk_3)+G(\CLbfk_1,\CLbfk_4)+
    G(\CLbfk_3,\CLbfk_4)- \notag\\ &&&
    -G(\CLbfk_1,\CLbfk_2+\CLbfk_4)-G(\CLbfk_3,\CLbfk_2+\CLbfk_4)-
    G(\CLbfk_2,\CLbfk_1+\CLbfk_3) - \notag\\ &&&
    -G(\CLbfk_4,\CLbfk_1+\CLbfk_3)+G(\CLbfk_1+\CLbfk_3,\CLbfk_2+\CLbfk_4)
    \bigr] + \notag \\ &
  +\delta^{ad}\delta^{bc} && \bigl[
    G(\bm k_1,\bm k_3)+G(\bm k_3,\bm k_4)+G(\bm k_1,\bm k_2)+G(\bm k_2,\bm k_4)
    - \notag\\ &&&
    -G(\CLbfk_1,\CLbfk_2+\CLbfk_3)-G(\CLbfk_4,\CLbfk_2+\CLbfk_3)-
    G(\CLbfk_3,\CLbfk_1+\CLbfk_4) - \notag\\ &&&
    -G(\CLbfk_2,\CLbfk_1+\CLbfk_4)+G(\CLbfk_1+\CLbfk_4,\CLbfk_2+\CLbfk_3)
    \bigr]
  \Bigr\} \,.
\end{alignat}
The convolution $D_2\otimes G$ is given by the expression \cite{BW}
\begin{align}
  (D_2\otimes G)(\CLbfa,\CLbfb)= 3g^2\int\!\!\frac{\CLrd^2k}{(2\pi)^3}
  \Biggl\{ & \Biggl[ \frac{\CLbfa^2}{(\CLbfk-\CLbfa)^2\CLbfk^2}+
      \frac{\CLbfb^2}{(\CLbfk+\CLbfb)^2\CLbfk^2}
      -\frac{(\CLbfa+\CLbfb)^2}{(\CLbfk-\CLbfa)^2(\CLbfk+\CLbfb)^2} 
      \Biggr] D_2(\CLbfk^2) - \notag\\
    & -\frac1{(\CLbfk-\CLbfa)^2}\Biggl[\frac{\CLbfa^2}{(\CLbfk-\CLbfa)^2+
      \CLbfk^2}-\frac{(\CLbfa+\CLbfb)^2}{(\CLbfk-\CLbfa)^2+(\CLbfk+\CLbfb)^2} 
    \Biggr] D_2(\CLbfa^2) -
      \notag\\
    & -\frac1{(\CLbfk+\CLbfb)^2}\Biggl[\frac{\CLbfb^2}{(\CLbfk+\CLbfb)^2+
      \CLbfk^2}-\frac{(\CLbfa+\CLbfb)^2}{(\CLbfk-\CLbfa)^2+(\CLbfk+\CLbfb)^2}
    \Biggr] D_2(\CLbfb^2)
  \Biggr\} \, .
\end{align}
For the twist analysis in DLA it is useful to perform a Mellin-transformation
in the following way (cf.\ eqn.\,\eqref{DoubleMelTrans1}):
\begin{equation}
  \bigl(D_2\otimes G\bigr)(\omega,\CLbfa,\CLbfb)=
  \int\!\! \frac{\CLrd\nu}{2\pi\CLri} \,\,
  D_2(\omega,\nu)\tilde{G}(\nu,\CLbfa,\CLbfb)
\end{equation}
with \cite{BW}
\begin{alignat}{2} \label{V_G_nuRep1}
  \tilde{G}(\nu,\bm a,\bm b)=\frac{3g^2}{(2\pi)^2}\Theta\bigl(|\bm a|-|\bm b|)
    \Biggl\{&
    \left[\ln\left(\frac{|\bm a|}{|\bm b|}\right)-\frac1\nu\right]
      \left(b^2\right)^{-\nu} &&+ \\ &
    +\sum_{n=1}^\infty\left(-\frac{|\bm b|}{|\bm a|}\right)^n\cos(n\theta)
      &&\biggl[\left(\frac1{n+\nu}-\frac1n\right)\left(a^2\right)^{-\nu}+
      \notag\\ &&&
      +\left(\frac1{n-\nu}-\frac1n\right)\left(b^2\right)^{-\nu}
      \biggr]
  \Biggr\} +\Bigl(|\bm a|\leftrightarrow |\bm b|\Bigr) \notag \, ,
\end{alignat}
where $\theta$ represents the angle between $\bm a$ and $\bm b$. 
The leading-twist part in the DLA can be obtained by evaluating the residue at
$\nu=-1$:
\begin{equation} \label{GnuRepT2}
  \tilde{G}(\nu,\bm a,\bm b)^{\tau=2}=-\frac{3g^2}{(2\pi)^2}
  \,\bm a\cdot\bm b\, 
  \frac{\left(\max\left(a^2,b^2\right)\right)^{-\nu-1}}{\nu+1} \,,
\end{equation}
whereas the twist-four contribution is related to the pole at $\nu=-2$:
\begin{equation} \label{GnuRepT4}
  \tilde{G}(\nu,\bm a,\bm b)^{\tau=4}=\frac{3g^2}{(2\pi)^2}
  \left[2(\bm a\cdot\bm b)^2-a^2\,b^2\right]
  \frac{\left(\max\left(a^2,b^2\right)\right)^{-\nu-2}}{\nu+2}\, .
\end{equation}

Inserting the twist-two result \eqref{GnuRepT2} in \eqref{Vdef1}, we find that
the sum of the various terms cancel. Returning to \eqref{Bet_Salp1}, this
cancellation means that we are loosing one $\ln Q^2$, i.e.\ we have a
non-leading correction to the DLA which is beyond our control and will
therefore be disregarded. For the twist-four result, on the other hand, we find
a nonzero coefficient of the pole near $\nu=-2$. In \eqref{Bet_Salp1} the
twist-four part of the first term on the r.h.s.\ reads (we use DLA accuracy,
and we assume, for simplicity, that all $k_i^2$ are of the same order, $k^2$):
\begin{align} \label{D2convVDLA1}
  \frac1\omega D_2^t \otimes V(\omega;\bm{k}_1,\bm{k}_2,\bm{k}_3,\bm{k}_4
  )^{\tau=4} = a_2^t \sum_f e_f^2 \frac{\sqrt8}{2\pi}
  2\sqrt2 \alphaS^3
  \frac{Q^2}{Q_0^2} \,
  \int\!\! \frac{\CLrd \tilde\nu}{2\pi\mathrm{i}}
  & \left( \frac1{k^2} \right)^{\tilde\nu} \,
  \frac{1}{\omega\tilde\nu}
  \frac{1}{(\omega\tilde\nu-\gamma_2)} \times \notag\\
  \times\bigl\{
    (-\delta^{ab}\delta^{cd}+\delta^{ac}\delta^{bd}+\delta^{ad}\delta^{bc})
      &\bm{k}_1\cdot\bm{k}_2\,\bm{k}_3\cdot\bm{k}_4 + \notag\\
  +\, (\delta^{ab}\delta^{cd}-\delta^{ac}\delta^{bd}+\delta^{ad}\delta^{bc})
      &\bm{k}_1\cdot\bm{k}_3\,\bm{k}_2\cdot\bm{k}_4 + \\
  +\, (\delta^{ab}\delta^{cd}+\delta^{ac}\delta^{bd}-\delta^{ad}\delta^{bc})
      &\bm{k}_1\cdot\bm{k}_4\,\bm{k}_2\cdot\bm{k}_3
  \,\bigr\} \, .\notag
\end{align}
Here $\tilde\nu\equiv\nu+2$, and all momenta are normalized by $Q$. The factors
in the first line of eqn.\,\eqref{D2convVDLA1} arise from the two-gluon
amplitude and the quark-loop in DLA, while the lower three lines correspond to
the transition $2\to 4$ gluons. A similar result holds for the longitudinal
structure function (with $a_2^t$ replaced by $b_2^{\ell}$, and an extra factor
$1/\tilde\nu$ in the first line).

Making systematic use of the double logarithmic approximation, it is possible
to find an explicit solution of eqn.\,\eqref{Bet_Salp1}.  To this end we solve
the integral equation by iteration: $\DfourI$ then represents a sum of diagrams
which have the following structure (fig.\,\ref{G_Lad1}b). At the top we start
with the quark-loop $D_{2,0}$ coupled to a BFKL ladder. At the lower end of
this ladder we have the $2\to4$ transition vertex $V$, then further below the
four-gluon state with the sum over all pairwise interactions. It is, now,
convenient to use the method of Faddeev \cite{Fad1}. We reorganize the sum over
all pairwise gluon interactions and introduce auxiliary potentials
$T_{(ij)(kl)}$ which describe pairs of two-gluon ladders between the gluon
lines $(ij)$ and $(kl)$. The original sum over all pairwise interactions
between the four gluon lines then translates into the iteration of these
auxiliary potentials, allowing for all possible recombinations (``switches'')
$(ij)(kl)\to (ik)(jl)$ etc. For each gluon ladder there are six irreducible
color states $(8 \otimes 8 = 1 \oplus 8_\mathrm{A} \oplus 8_\mathrm{S} \oplus
10 \oplus \bar{10} \oplus 27)$. However, if we add the first rung right below
the vertex in eqn.\,\eqref{D2convVDLA1}, e.g.\ between the gluons 1 and 2, and
extract the leading power of $\ln (k^2/Q^2)$ the three terms in
\eqref{D2convVDLA1} become proportional to $\bm k_1\cdot\bm k_2\bm k_3\cdot\bm
k_4$ and the anti-symmetric color states cancel, so that only the symmetric
ones ($1$, $8_\mathrm{S}$, $27$) survive. This remains true if one adds further
rungs. As a result, our four-gluon system has only nine components: three color
components for each coupling scheme $(12)(34)$, $(13)(24)$ and $(14)(23)$.
Concentrating, now, on the twist-four point $\nu=-2$ and collecting the maximum
number of logarithms, we find the following pattern. For each two-ladder state
described by the auxiliary potential $T_{(ij)(kl)}$ we have a propagator
$\gamma_i$:
\begin{equation} \label{gammaEleDef1}
  \gamma_i = \left( 1-\frac{4\sigma_i}{\omega\tilde\nu}\right)^{-\frac12}-1 \,,
\end{equation}
where $\pi\sigma_i/\alphaS=\{3,3/2,-1\}$ for the color states $i=\{1,8_\mathrm
S,27\}$. At each switch from the two-ladder state $(ij)(kl)$ to the
configuration $(ik)(jl)$, we collect a factor $1/2$ from the angular
integration, and insert a color recoupling matrix $S$. In our nine-component
matrix notation, the Green's function $\Sigma$ for the four-gluon system then
becomes:
\begin{equation} \label{SigmaDef1}
  \Sigma = \sum\limits_{n=0}^\infty\left( G\frac12 S\right)^n G =
  \left(G^{-1}-\frac12 S\right)^{-1}, \quad
  G=\begin{pmatrix}
    \gamma & 0 & 0 \\
    0 & \gamma & 0 \\
    0 & 0 & \gamma
    \end{pmatrix}\, , \quad
  S=\begin{pmatrix}
    0 & \Lambda & \Lambda \\
    \Lambda & 0 & \Lambda \\
    \Lambda & \Lambda & 0
    \end{pmatrix} \, \, ,
\end{equation}
where the sub-matrices $\gamma$ and $\Lambda$ have the following components:
\begin{equation} \label{gam_and_lam_def1}
  \gamma=\begin{pmatrix}
    \left(1-\frac{4\sigma_1}{\omega\tilde\nu}\right)^{-\frac12} -1 & 0 & 0 \\
    0 & 
    \left(1-\frac{4\sigma_{8_\mathrm{S}}}{\omega\tilde\nu}\right)^{-\frac12}-1
      & 0 \\
    0 & 0 & \left(1-\frac{4\sigma_{27}}{\omega\tilde\nu}\right)^{-\frac12}-1
  \end{pmatrix} \qquad
  \Lambda=\begin{pmatrix}
    \frac18 & \frac{1}{2\sqrt{2}} & \frac{3\sqrt{3}}{8} \\
    \frac{1}{2\sqrt{2}} & -\frac{3}{10} & \frac{3}{10}\sqrt{\frac32} \\
    \frac{3\sqrt{3}}{8} & \frac{3}{10}\sqrt{\frac32} & \frac{7}{40}
  \end{pmatrix} \,\, .
\end{equation}
As a result of the leading-$\ln Q^2$ approximation, at the lower end of the
Green's function $\Sigma$ the two ladder system $T_{(ij)(kl)}$ ends up with the
momentum configuration $\CLbfk_i=-\CLbfk_j\equiv \CLbfl$ and
$\CLbfk_k=-\CLbfk_l\equiv \CLbfl'$, and for simplicity we set
$\CLbfl^2={\CLbfl'}^2=Q_0^2/Q^2$.

For the coupling of this Green's function to the transition vertex we introduce
the nine component vector $\CLbfV$: $\CLbfV^\mathrm T=(\CLbfv^\mathrm
T,\CLbfv^\mathrm T,\CLbfv^\mathrm T)$ with
\begin{equation} \label{Vcomp1}
  \CLbfv^\mathrm T = (2,\, 4\sqrt{2},\, 6\sqrt{3}) \,.
\end{equation}
This vector is obtained from \eqref{D2convVDLA1} by applying the color
projectors listed in the appendix. As a result, our function $\DfourI$, written
as a (transposed) nine component vector, takes the form:
\begin{eqnarray} \label{D4_eqn1}
  \bm D_4^{\mathrm I,t}(\omega,Q_0^2/Q^2) &= &
  \frac1{\omega} (D_2^t \otimes \CLbfV^\mathrm{T})(\omega,Q^2)^{\tau=4}
  + \\ &&+\,
  a_2^t \sum\limits_f e_f^2 \frac{\sqrt8}{2\pi}
  2\sqrt2 \alphaS^3
  \left( \frac{Q_0^2}{Q^2} \right)^2 \,
  \int\!\! \frac{\CLrd \tilde\nu}{2\pi\mathrm{i}}
  \left( \frac{Q^2}{Q_0^2} \right)^{\tilde\nu} \,
  \frac{1}{\omega\tilde\nu}
  \frac{1}{(\omega\tilde\nu-\gamma_2)} \cdot
  \CLbfV^\mathrm{T}\, \Sigma(\omega\tilde\nu)\, \,\, . \nonumber
\end{eqnarray}

Finally we have to couple the four-gluon system to the proton. In modelling
this coupling we take, as a guideline, the structure of $D_4$: to be definite,
we assume that the coupling to the proton has the same symmetry properties as
$D_4$ in \eqref{D4sumD4R_D4I}, the coupling of the four-gluon state to a
perturbative target, e.g.\ to a virtual photon or to a heavy onium state. The
decomposition in \eqref{D4sumD4R_D4I} reflects the different contributions:
apart from totally symmetric (under the combined interchange of color and
momenta) terms which are contained in $\DfourI$ and $\DfourR$, we have a term
with mixed symmetry properties which occurs only in $\DfourR$. For the
symmetric piece we make the simplest ansatz, which contains only the color
singlet representation.  For the term with the mixed symmetry we simply follow
the perturbative ansatz. This leads to:
\begin{equation} \label{initialtotal}
  \varphi_4^{abcd}= \varphi_{4\CLrS}^{abcd} + \varphi_{4\CLrA}^{abcd}
\end{equation}
with
\begin{equation} \label{initialsym}
  \varphi_{4\CLrS}^{abcd}=\frac{1}{3\cdot8} \frac{1}{k_1^2k_2^2k_3^2k_4^2}
    \left(
      \delta^{ab}\delta^{cd} f_\CLrS(1,2;3,4;\omega)+
      \delta^{ac}\delta^{bd} f_\CLrS(1,3;2,4;\omega)+
      \delta^{ad}\delta^{bc} f_\CLrS(1,4;2,3;\omega)   \right)
\end{equation}
and 
\begin{multline} \label{initialasym}
  \varphi_{4\CLrA}^{abcd}=-\frac{1}{3\cdot8} \frac{1}{k_1^2k_2^2k_3^2k_4^2}
    \Bigl( f^{abm}f^{mcd} f_\CLrA(1,2;3,4;\omega)+ \\
      + f^{acm}f^{mbd} f_\CLrA(1,3;2,4;\omega)+
      f^{adm}f^{mbc} f_\CLrA(1,4;2,3;\omega) \Bigr)
\end{multline}
with positive-valued functions $f_\CLrS$ and $f_\CLrA$. The signs follow from
the AGK rules and will be discussed below in more detail. The pre-factors 1/8
and 1/3 are convenient for color and for statistics, resp.  We immediately see
that the $\varphi_{4\CLrA}$-term does not couple to $\DfourI$ but only to
$\DfourR$, and we will come back to this term when we discuss $\DfourR$. In
order to discuss the momentum structure of the ansatz $\varphi_{4\CLrS}$, let
us recall that in the usual DIS leading-twist ladder the lowest cell has only a
longitudinal integration, and the integration over the virtuality of the lowest
parton is absorbed into the initial condition $\varphi_2(Q_0^2,x)$.  In
evaluating our four-gluon system in the DLA we treat each gluon ladder of our
auxiliary potential $T_{(ij)(kl)}$ in the same way as the leading-twist ladder.
Let the lowest potential be $T_{(12)(34)}$.  At its lower end we have the
momentum factor $(\bm k_1 \bm k_2)(\bm k_3\bm k_4)$.  In the function
$f_\CLrS(i,j;k,l;\omega)$ of our ansatz \eqref{initialtotal} the main
contribution comes from the region where
\begin{align} \label{correl}
  \bm k_i &= \bm l+\bm r, \quad\bm k_j=-\bm l+\bm r, \quad
  \bm k_k = \bm m-\bm r,\quad \bm k_l=-\bm m-\bm r, \nonumber \\
       r^2 &\ll l^2,\, m^2 \,.
\end{align}
Combining these functions with the momentum factors from the auxiliary
potential $T_{(12)(34)}$ we are lead to the following definition of our initial
condition:
\begin{equation} \label{initialinter}
  Q_0^2 \, \varphi_{4\CLrS}(Q_0^2;\omega) = \int^{Q_0^2}\!\!\CLrd r^2 
    \int_{r^2}^{Q_0^2}\!\frac{\CLrd l^2}{l^4}
    \int_{r^2}^{Q_0^2}\!\frac{\CLrd m^2}{m^4} l^2m^2
    f_\CLrS(k_i,k_j;k_k,k_l;\omega).  
\end{equation}
Inserting into \eqref{initialsym} the momentum assignment \eqref{correl}, and
using the definition \eqref{initialinter} we get a factor
\begin{equation} \label{initialfinal}
  \frac{1}{3\cdot8} \left( \delta^{ab}\delta^{cd} + \frac{1}{2}
    \delta^{ac}\delta^{bd}
    +\frac{1}{2} \delta^{ad}\delta^{bc}\right) 
      Q_0^2 \cdot \varphi_{4\CLrS}(Q_0^^2;\omega),
\end{equation}
where the factors 1/2 in front of the second and the third term are due to the
angular integrations over $\bm l$ and $\bm m$ (a similar definition holds for
the functions $\varphi_{4\CLrA}$ and $f_\CLrA$). As to the longitudinal
integrals, since we work in $\omega$-space we allow the function
$\varphi_{4\CLrS}$ to depend on $\omega$: if we assume a single ladder to rise
as $(1/x)^{\lambda}$, $\varphi_{4\CLrS}$ should have a pole at $\omega = 2
\lambda$, i.e.\ the initial distribution rises twice as strong as the
leading-twist gluon structure function at the same scale.

Finally we couple this ansatz to $\DfourI$ in \eqref{D4_eqn1}. Projecting onto
the color eigenstates and introducing the nine-component vector
\begin{equation} \label{initialvector}
  {\bm \Phi}_{4\CLrS}(Q_0^2,\omega) =
  \begin{pmatrix}
    \bm{w}\\
    \bm{w}\\
    \bm{w}
  \end{pmatrix} 
  \cdot \varphi_{4\CLrS}(Q_0^2;\omega), \qquad
  \bm w=\frac{1}{3\cdot8}
  \begin{pmatrix}
    9\\
    2\sqrt{2}\\
    3\sqrt{3}
  \end{pmatrix}
\end{equation}
we arrive at the following expression for our twist-four contribution to the
structure function in the $(\omega,\nu)$-representation:
\begin{eqnarray} \label{D4final}
  \Delta F_t^\mathrm I & = & -\frac{1}{128 \pi^2} \frac{Q^2}{Q_0^2} 
  D_4^{\mathrm I;abcd}(\omega,\nu) \varphi_4^{abcd}(\omega) =
    \notag \\ &=&
    -\frac{a_2^t}{16} \frac{\alphaS^3}{\pi^3}
    \sum_f e_f^2 \frac{Q_0^2}{Q^2} \,
    \int\!\! \frac{\CLrd \tilde\nu}{2\pi\mathrm{i}}
    \left( \frac{Q^2}{Q_0^2} \right)^{\tilde\nu} \,
    \frac{1}{\omega\tilde\nu}
    \frac{1}{(\omega\tilde\nu-\gamma_2)} 
    \varphi_{4\CLrS}(\omega)
  \\ \notag
    && -\, \frac{a_2^t}{32} \frac{\alphaS^3}{\pi^3}
    \sum\limits_f e_f^2 \frac{Q_0^2}{Q^2} \,
    \int\!\! \frac{\CLrd \tilde\nu}{2\pi\mathrm{i}}
    \left( \frac{Q^2}{Q_0^2} \right)^{\tilde\nu} \,
    \frac{1}{\omega\tilde\nu}
    \frac{1}{(\omega\tilde\nu-\gamma_2)} \cdot
    \CLbfV^\mathrm{T}\, \Sigma(\omega\tilde\nu)\, \bm \Phi_{4\CLrS} (\omega),
\end{eqnarray}
and an analogous expression in the longitudinal case. After performing the
inversion within $\Sigma=(G^{-1}-\frac12S)^{-1}$ we get explicit expressions
for the components of vector $\Sigma\bm V$. They are listed in the appendix.

As one of the most striking features of (\ref{D4final}) we note that $\Sigma$
has a leading singularity in the $\tilde\nu$-plane at \cite{B,LRS}
\begin{eqnarray} \label{Pol_Pos1}
  \tilde\nu= 4(1+ \delta) \frac{N_\mathrm c\alphaS}{\pi \omega} \,\, ,
\end{eqnarray}
where $\delta=0.009549$. Using the usual saddle point approximation, this pole
leads to the following asymptotic small $x$ behavior:
\begin{equation}
  \DfourI \sim \left( \frac{Q_0^2}{Q^2} \right)^2 \exp \left(2
    \left(1+\frac{\delta}{2} \right)
    \sqrt{\frac{4N_c}{\pi} \alphaS \ln (1/x) \ln (Q^2/Q_0^2)} \right) \,.
\end{equation}
This rise at small $x$ is approximately twice as strong as that of the
leading-twist gluon structure function, and at sufficiently small $x$ the
four-gluon higher-twist term will become as strong as the leading-twist term,
despite its suppression coming from the extra power in $\alphaS$ and the
$1/Q^2$ factor. This strong rise at small $x$ is the reason why the four-gluon
operator, \eqref{Four_G_Op1}, is expected to be so important in the small $x$,
low $Q^2$-region.

Besides the leading pole \eqref{Pol_Pos1}, there are more singularities in the
$\tilde{\nu}$-plane, which have been discussed in some detail in ref.
\cite{BR1}. There are three cuts in the $\tilde\nu$-plane. The first one stems
from the singlet-part of matrix $\gamma$ (eqn.\,\eqref{gam_and_lam_def1}). It
starts at $\tilde\nu=0$ and goes up to the point $\tilde\nu= 4 \frac{N_\mathrm
  c\alphaS}{\pi \omega}$ which is very close to the leading pole
\eqref{Pol_Pos1}. It overlaps with the symmetric color octet cut, which is
located in the interval $[0,\frac{6\alphaS}{\pi\omega}]$. Finally, the cut of
the color 27plet goes from $-\frac{4\alphaS}{\pi\omega}$ to $0$.
\begin{figure}
  \begin{center}
    \input{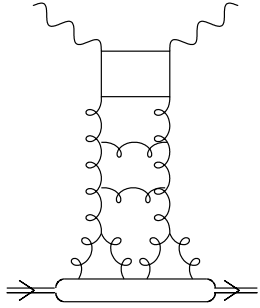}
    \caption{The reggeizing part of the four-gluon amplitude has the same
      evolution as the two-gluon amplitude but is suppressed by one
      factor $\alphaS$.} \label{D4Rfig1}
  \end{center}
\end{figure}
%

Finally, we return to \eqref{D4sumD4R_D4I} and we discuss the {\it reducible}
part $\DfourR$. From \cite{B,BW} we have:
\begin{align} \label{D4R}
  \DfourR(\CLbfk_1,\CLbfk_2,\CLbfk_3,&\CLbfk_4) = &&
  \\ 
  =\frac{g^2}{2\sqrt{2}} \Bigl\{ \phantom{+}
    &d^{abcd} \bigl[D_2(\CLbfk_1,\CLbfk_2+\CLbfk_3+\CLbfk_4)+
      D_2(\CLbfk_4,\CLbfk_1+\CLbfk_2+\CLbfk_3) &\!\!\!\!\!\!-\,\,&
      D_2(\CLbfk_1+\CLbfk_4,\CLbfk_2+\CLbfk_3) \bigr] +
  \nonumber \\ 
    +\, & d^{abdc} \bigl[
      D_2(\CLbfk_2,\CLbfk_1+\CLbfk_3+\CLbfk_4) +
      D_2(\CLbfk_3,\CLbfk_1+\CLbfk_2+\CLbfk_4) &\!\!\!\!\!\!-\,\,&
      D_2(\CLbfk_1+\CLbfk_2,\CLbfk_3+\CLbfk_4) -
  \notag \\ & \notag
      &\!\!\!\!\!\!-\,\,& D_2(\CLbfk_1+\CLbfk_3,\CLbfk_2+\CLbfk_4) \bigr]\, 
      \Bigr\} \,,
\end{align}
where the color tensor $d^{abcd}$ is given in the appendix. It contains both
symmetric pieces and pieces with mixed symmetry: one easily recognizes the
structures \eqref{initialsym} and \eqref{initialasym}. Convoluting $\DfourR$
with our ansatz \eqref{initialtotal} we find that $\varphi_{4\CLrS}$ couples
only to the symmetric and $\varphi_{4\CLrA}$ only to the mixed-symmetric part
of $\DfourR$. With the help of the same arguments as before,
cf.\,\eqref{initialinter}, we arrive at
\begin{align} \label{deltaFR}
  \Delta F_t^\mathrm R &= -\frac{1}{128\omega \pi^2}
  \left(\frac{Q^2}{Q_0^2}\right) 
  D_4^{\mathrm R;abcd}\, \varphi_4^{abcd}(\omega)^{\tau=4} =\notag \\
  &= \frac{a_2^t}{64}\frac{\alphaS^2}{\pi^2}
  \sum_f e_f^2\, \frac1\omega  \left( \frac{Q_0^2}{Q^2} \right)
  \exp\left(\frac{\gamma_2}{\omega}\ln(Q^2/Q_0^2)\right)
    \frac{1}{3} [14\varphi_{4\CLrS}(\omega)- 9\varphi_{4\CLrA}(\omega)] \,.
\end{align}
Here $\varphi_{4\CLrA}$ is defined in analogy with $\varphi_{4\CLrS}$ in
\eqref{initialinter}. It is important to note that the negative sign in front
of $\varphi_{4\CLrA}$ is dictated by the AGK rules (see below).  Note that, to
lowest order in $\alphaS$, it is this contribution $\DfourR$ that contains the
diagrams with four elementary gluon lines between the quark-loop and the proton
(fig.\,\ref{FGLeadOrd1}).

Having written down our DLA expressions for the four-gluon twist-four
contributions, we briefly return to the operator product expansion. As before,
the ladder diagrams denote the $\ln Q^2$ evolution in the DLA, but instead of
writing the four-gluon amplitude as an evolution equation in $\ln Q^2$ we have
used the closed expression in the $(\omega,\nu)$-representation. Twist-four
corresponds to the point $\nu=-2$.  $\DfourI$ then has the following
interpretation: the upper two-gluon ladder (evaluated near $\nu=-2$) describes
the evolution of operator \eqref{Two_G_Op1}, the $2\to4$ vertex the mixing
between \eqref{Four_G_Op1} and \eqref{Two_G_Op1}, and the Green's function
$\Sigma$ the evolution of the four-gluon operators \eqref{Four_G_Op1}.
Obviously, the three auxiliary potentials $T_{(12)(34)}$, $T_{(13)(24)}$,
$T_{(14)(23)}$ belong to the three operators in the first, second, and third
line of \eqref{Four_G_Op1}, and what we have called ``switching from $(ij)(kl)$
to $(ik)(jl)$'' has to be interpreted as the mixing between these three
different four-gluon operators. If we were to go beyond DLA, there would also
be a transition from the four-gluon state back to two gluons (this vertex could
again be derived from \eqref{Vdef1} and \eqref{V_G_nuRep1}, by expanding about
$\nu=1$), and eventually these transitions would be iterated and exponentiated
in the $t$-channel. A remarkable feature of the small $x$ approximation (upon
which the DLA is based) is the fact that the four-gluon operator has no direct
coupling to the quark-loop. Our ansatz \eqref{initialtotal} for the coupling to
the proton, i.e.\ for the initial condition, which we so far have motivated by
inspecting the structure of $D_4$, also arises from an analysis in the spirit
of \cite{EFP}: the matrix element of the product of four-gluon fields has to be
decomposed into terms with different symmetry properties.

The interpretation of the contribution $\DfourR$, on the other hand, is more
subtle. As stated before, this term has its origin in the reggeization of the
gluon: from the point of view of DIS the appearance of reggeization is a
``novel'' phenomenon since at the leading-twist level it has played only an
indirect r\^ole (e.g.\ as part of the NLO anomalous dimension of the
gluon). From the point of view of $Q^2$ evolution, $\DfourR$ belongs to the
two-gluon operator \eqref{Two_G_Op1}. Including $\DfourR$, therefore, at first
sight seems like adding a non-leading correction to \eqref{D2final}, and within
the DLA we better should restrict ourselves to the leading term
\eqref{D2final}. On the other hand, the analysis in \cite{BW} clearly shows
that $\DfourI$ and $\DfourR$ have the same origin (diagrams with four gluons in
the $t$-channel) and come with the same powers of logarithms. We, therefore,
believe that for the twist-four calculations -- which necessarily have to
include $t$-channel states with four gluons -- we have to keep $\DfourR$ as a
separate contribution, in addition to \eqref{D2final}.

A better understanding of why we have to keep this contribution can be obtained
by looking at the evolution of the gluonic operator in \eqref{Two_G_Op1}. The
mixing between \eqref{Four_G_Op1} and \eqref{Two_G_Op1} is of order $\alphaS^2$
in the anomalous dimension matrix, i.e.\ in total the correction to the
evolution of \eqref{Two_G_Op1} due to the transition from \eqref{Two_G_Op1} to
\eqref{Four_G_Op1} and back to \eqref{Two_G_Op1} is a NNLO effect of order
$\alphaS^3$. To be consistent, we, therefore, have to include NNLO corrections
also in the anomalous dimension $\gamma^{\tau=4}$ in \eqref{BFKLanomDims1} and
in the coefficient function. They are not available yet. On the other hand,
what we are seeing in $\DfourR$ can be viewed as a particular piece of these
corrections, connected with the reggeization of the gluon. Moreover, there are
reasons to expect that they are the most important ones. Since we are studying
$t$-channel states with four gluons, $\DfourR$ provides contributions from the
``decay of a reggeizing gluon into two or three gluons'' which then become
partons inside the proton. In other words, we are encountering the two- and
three-particle states in the gluon trajectory function, appearing at the lower
end of the two-gluon ladder. Clearly, these higher-order particle states in the
trajectory function can appear not only at the lower end of the two-gluon
ladder (as taken into account by $\DfourR$) but also somewhere between the
proton and the photon. Such a contribution would then be part of the NNLO
corrections to the gluon rung, i.e.\ to the anomalous dimension
$\gamma^{\tau=4}$ in \eqref{BFKLanomDims1} mentioned above. In this sense,
$\DfourR$ counts those NNLO corrections to the two-gluon kernel which are
connected with four-gluon states in the $t$-channel, and these corrections are
taken into account only at the lower end of the ladder. When running $\alphaS$
and the strong ordering of the transverse momenta is taken into account, we
expect these corrections to be largest when they appear at the lower end of the
ladder.  For these reasons our incomplete treatment of the NNLO corrections in
the evolution of the two-gluon operator \eqref{Two_G_Op1} may not be such a bad
approximation.
%
%
\subsection{Contributions with Three $t$-channel Gluons}
Since there are no three-gluon operators that might contribute to our
twist-four analysis of the unpolarized structure functions, $t$-channel states
with three gluons can come in only through the mechanism which we have
discussed at the end of the previous section. In the evolution equations of the
two-gluon operator \eqref{Two_G_Op1} we should include the NLO corrections to
the anomalous dimension $\gamma^{\tau=4}$ in \eqref{BFKLanomDims1} as well as
to the coefficient function. Whereas the former ones are, at least in
principle, now available (so far, they have not been computed from \cite{FL};
as pointed out in \cite{Salam}, it is not clear how the behavior near $\nu =
-1$ can directly be extracted from the results of \cite{FL}), NLO corrections
to the coefficient function are not known. In our analysis, therefore, we
follow the same logic as for the four-gluon case. From \cite{BW} we know
scattering amplitudes with three gluons in the $t$-channel (in \cite{BW} they
are called $D_3$). Because of the reggeization of the gluon they can also be
written as a sum of $D_2$ functions (cf.\,\eqref{D4R}):
\begin{equation} \label{D3Rel1}
  D_3(\CLbfk_1,\CLbfk_2,\CLbfk_3)=\frac{g}{4\sqrt{2}}f^{abc}\left(
    D_2(\CLbfk_1+\CLbfk_2,\CLbfk_3) + D_2(\CLbfk_1,\CLbfk_2+\CLbfk_3)-
    D_2(\CLbfk_1+\CLbfk_3,\CLbfk_2) \right).
\end{equation}
The convolution with the initial distribution of the proton requires
a new function, $\varphi_3(\omega)$. One ends up with functions 
$\Delta F_{(t,\ell)}^3$ that have the same dependence on variables as
\eqref{deltaFR} (with a different constant factor in front of the r.h.s.;
it will be absorbed into the unknown initial distribution $\varphi_3$).
%
%
\subsection{Running $\alphaS$}
Since in DLA the leading-twist two-gluon amplitude must coincide with the DGLAP
solution, it is clear how one can consider the fact that $\alphaS$ is running
in this case. One has to exchange the value of fixed $\alphaS$ by
$\bar\alphaS\equiv 4\pi/\beta_0$ ($\gamma_2\to 12/\beta_0$) and the expression
$Q^2/Q_0^2$ by $t/t_0$, where $t=\ln (Q^2/\Lambda^2)$ and $t_0=\ln
(Q_0^2/\Lambda^2)$. This procedure should be applicable for the bare four-gluon
amplitude (i.e.\ the amplitude without two-gluon amplitude and vertex at the
top of fig.\,\ref{G_Lad1}b), as well, since the addition of one rung is always
performed by convolution with the common BFKL kernel. This means that every
time $\alphaS$ comes with a factor $\ln(t/t_0)$ it has to be replaced by
$\bar\alphaS$.

There are three powers of $\alphaS$ in eqn.\,\eqref{D4final}: One stems from
the quark-loop, which (in the longitudinal case at leading order) is
proportional to a logarithm. The remaining two powers stem from the coupling of
the four-gluon to the two-gluon amplitude via the Vertex $V$
(fig.\,\ref{G_Lad1}b). Here we get only one logarithm. We write this coupling
symbolically as
\begin{equation} \label{CoupInt1}
  D_4^{\mathrm I,\ell}=
  \int_{\tau_0}^{\tau}\!\! \CLrd\tau'
  \exp[\gamma_2(\tau-\tau')] f \alphaS^2
  \int\!\! \frac{\CLrd \nu}{2\pi\mathrm{i}}
  \exp[\nu(\tau'-\tau_0)] G(\nu) \,\, ,
\end{equation}
where $\tau=\ln t$ etc., $G(\nu)$ is understood as the \textit{four-gluon
  amplitude Greens-function} and $f$ shall contain all remaining factors.  In
eqn.\,\eqref{CoupInt1} one factor $\alphaS$ needs to be replaced by $\alphaS
(t')$ and the other one by $\bar\alphaS$. If, on the other hand, we go back to
fixed $\alphaS$, we have to replace $\tau$ and $\tau'$ in
eqn.\,\eqref{CoupInt1} by $t$ and $t'$, resp. Performing the $t'$-integration
and comparing with eqn.\,\eqref{D4final}, we can identify the quantities $G$
and $f$. By this way we arrive at the following conclusions: For running
$\alphaS$ the factor $(Q^2/Q_0^2)^\nu$ in eqn.\,\eqref{D4final} has to be
replaced by $(t/t_0)^\nu$, one power $\alphaS$ has to be evaluated as
$\alphaS(Q^2)$, while (in the longitudinal case) the remaining two powers are
replaced by $\bar\alphaS$. In the denominators we have to replace the factor
$(\omega\nu-\gamma_2)$ by $(\omega\nu-\omega-\gamma_2)$.

Repeating in (\ref{D4_eqn1}) the standard saddle point analysis
(cf.\ eqn.\,\eqref{Pol_Pos1}), we arrive at a small $x$ behavior of the form
\begin{eqnarray}
  \label{D4_Sad_Approx1}
  \DfourI \sim \left(\frac{Q_0^2}{Q^2}\right)^2 \,
  \exp 
  \left[ 
    2\left(1+\frac{\delta}{2}\right)
    \sqrt{\frac{48}{\beta_0}\, \ln\frac{1}{x}\,\ln\frac{t}{t_0}}
  \, \right] \,\, .
\end{eqnarray}
Apart from the small $\delta$-correction, this small $x$ behavior is just the
square of the well-known leading-twist double scaling formula \cite{DR,BF2}.
Therefore, in the kinematical regime in which $\DfourI$ gives the main
contribution to twist-four, higher-twist is expected to increase much faster
with decreasing $x$ than leading-twist and to decrease slower than $1/Q^2$.

We end this section with a brief summary. All in all we have collected four
different contributions (both for the transverse and the longitudinal
structure function). In our notation
\begin{equation} \label{DeltaFsumDs1}
  \Delta F = \DelFtwo + \DelFthree + \Delta F^\mathrm R +
  \Delta F^\mathrm I \,,
\end{equation}
where we have suppressed labels $t$ and $\ell$. As far as the $Q^2$-evolution
is concerned, the first three terms belong to the twist-four two-gluon operator
\eqref{Two_G_Op1} whereas the last one belongs to the four-gluon operators
\eqref{Four_G_Op1}. This last contribution is of particular interest, since at
small $x$ it rises stronger than the other ones and, therefore, may potentially
become large. However, we have already seen that the four contributions in
\eqref{DeltaFsumDs1} come with alternating signs, which hints at the
possibility of strong cancellations.
%
%
\begin{figure}
  \begin{center}
    \input{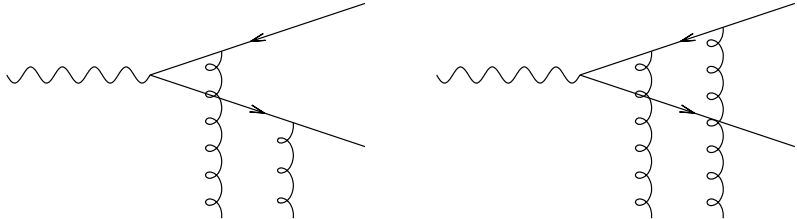}
    \caption{Two of the four diagrams which enter into diffractive
      $q\bar q$-production at leading order.}
    \label{qqbarFeyn1}
  \end{center}
\end{figure}
\begin{figure}
  \begin{center}
    \input{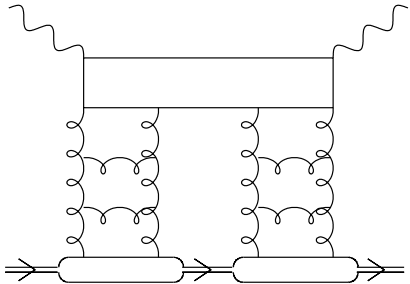}
    \caption{The twist-four part of diffractive $q\bar q$-production
      is described by exchange of gluon ladders.} \label{Diff_Lad1}
  \end{center}
\end{figure}
%
%
\section{Diffractive Dissociation} \label{SecDiffDiss}
Up to now, we have been describing the four contributions to twist-four, which
we believe to dominate at small $x$: $\DfourR$, $\DfourI$, $D_3$ and the
twist-four part of the two-gluon amplitude. Next, we have to address the
important question of specifying the initial distributions $\varphi$. Following
the conventional leading-twist analysis, we would determine the initial
conditions from a fit to the data. In this first attempt of analysing the
influence of twist-four corrections at low $Q^2$ and small $x$, we will try to
estimate the $\varphi$ functions with the help of empirical data on final
states which are known to belong to twist-four. The best candidates for this
are diffractive final states, in particular the production of longitudinal
vector particles and the production of jets with large transverse momenta. In
this section we, therefore, give a brief overview over the r\^ole of
higher-twist in DIS diffraction. In particular, we will discuss the relation
between the higher-twist pieces of diffractive cross sections and the
higher-twist contributions to $F_t$ and $F_{\ell}$ from the previous section.
In our numerical analysis these relationships will be used to estimate the
initial distribution $\varphi_{4\CLrS,\CLrA}$.
\begin{table}
  \begin{center}
    \begin{tabular}{l|ccc}
      Contribution & Sign & $\ln(Q^2/Q_0^2)$ & $\beta\to 1$ \\
      \hline
      Trans.\ $\tau=2$ & $+$ & non-pert. & vanishes \\
      Trans.\ $\tau=4$ & $-$ & no  & vanishes \\
      Long.\ $\tau=2$ & vanishes & vanishes & vanishes\\
      Long.\ $\tau=4$ & $+$ & yes & constant
    \end{tabular}
    \caption{The four contributions to diffractive $q\bar q$-production
      up to twist $\tau=4$ enter with different signs into the diffractive
      cross section. The table shows also which contributions have leading
      logarithms, and how they behave for $\beta\to 1$.}
    \label{SignAndLogsToDiff1}
  \end{center}
\end{table}
%
%
\subsection{Diffractive $q\bar q$-Production}
It is useful to divide the kinematical regime into different regions of the
variable $\beta\equiv Q^2/(M^2+Q^2)$, where $M$ is the invariant mass of the
diffractive system. We begin with the small mass region $(\beta\to 1)$, where
the $q\bar q$-production process is expected to give the main contribution to
diffraction. Those contributions to the cross section, which can be calculated
perturbatively, can be described as follows.  Seen from the angle of the proton
rest frame the virtual photon splits up into a $q\bar q$-pair, which interacts
with the proton before the quarks result into the two jets. For diffraction the
interaction between the $q\bar q$-pair and the proton must be colorless and,
therefore, in the small $x$ regime where gluons are expected to dominate, at
least two gluons need to be exchanged (fig.\,\ref{qqbarFeyn1}). The cross
section can, now, be computed by coupling the gluons in fig.\,\ref{qqbarFeyn1}
to the proton and by squaring the diagrams afterwards. This leads to the
following expressions in the case of zero momentum transfer $(t=0)$
\cite{BLW,NikoZak1}:
\begin{equation} \label{DiffCrossSecT1}
  \left. \frac{\CLrd\sigma^t}{\CLrd M^2\,\CLrd t\,\CLrd p_t^2}
  \right|_{t=0}
  = \sum_f e_f^2\frac{\alphaem\pi^2\alphaS^2}{12}\frac{1}{M^4}
  \frac{\left(1-\frac{2p_t^2}{M^2}\right)}{\sqrt{1-\frac{4p_t^2}{M^2}}} 
  \left[-\int\!\frac{\CLrd l^2}{l^2} \mathcal{F}_\mathrm{G}(\CLxP,l^2) \,
    \mathcal{I}^{\,t}(Q^2,M^2,p_t^2,l^2) \right]^2
\end{equation}
and
\begin{equation} \label{DiffCrossSecL1}
  \left. \frac{\CLrd\sigma^\ell}{\CLrd M^2\,\CLrd t\,\CLrd p_t^2}
  \right|_{t=0} =
  \sum_f e_f^2\frac{\alphaem\pi^2\alphaS^2}{3}\frac{4}{Q^2M^2}\frac{p_t^2}{M^2}
  \frac1{\sqrt{1-\frac{4p_t^2}{M^2}}} 
  \left[-\int\!\frac{\CLrd l^2}{l^2} \mathcal{F}_\mathrm{G}(\CLxP,l^2) \,
    \mathcal{I}^\ell(Q^2,M^2,p_t^2,l^2) \right]^2 \, .
\end{equation}
Here $\bm{p}_t$ is the transverse part (in the Sudakov decomposition) of the
quark-momenta, $\CLxP=x_\mathrm{B}/\beta$,
\begin{equation}
  \mathcal{I}^{\,t}(Q^2,M^2,p_t^2,l^2) = \frac{M^2-Q^2}{M^2+Q^2} +
  \frac{l^2 + \frac{p_t^2}{M^2}(Q^2-M^2)}
    {\sqrt{\left[l^2+\frac{p_t^2}{M^2}(Q^2-M^2)\right]^2 +
      4p_t^4\frac{Q^2}{M^2}}} \, ,
\end{equation}
\begin{equation}
  \mathcal{I}^\ell(Q^2,M^2,p_t^2,l^2) = \frac{Q^2}{M^2+Q^2} -
  \frac{p_t^2 Q^2}{M^2\sqrt{\left[l^2+\frac{p_t^2}{M^2}(Q^2-M^2)\right]^2 +
      4p_t^4\frac{Q^2}{M^2}}}
\end{equation}
and $\mathcal{F}_\mathrm{G}$ represents the unintegrated gluon distribution
of the proton
\begin{equation}
  \int^{Q^2} \!\!\CLrd l^2\, \mathcal{F}_\mathrm{G}(\CLxP,l^2) =
  \CLxP\, g(\CLxP,Q^2) \, .
\end{equation}

In order to establish a connection with the deep inelastic structure functions
we have to integrate over $M^2$ (or $\beta$), $p_t^2$ and $t$. Performing in
eqns.\,\eqref{DiffCrossSecT1} and \eqref{DiffCrossSecL1} the $p_t^2$- and
$M^2$-integrations one arrives at the quark-loop at the top of
fig.\,\ref{Diff_Lad1}, with the constraint that the two-gluon systems at the
r.h.s.\ and at the l.h.s.\ are in a color singlet state. We denote this with
the help of the superscript $(1,+,+)$: $D_{4,0}^{(1,+,+)}$. In terms of
$D_{4,0}^{(1,+,+)}$ the integrated diffractive cross section becomes:
\begin{equation} \label{qqbar}
  \left. \frac{\CLrd\sigma^{q\bar{q}}}{\CLrd t}\right|_{t=0}=
  \frac{4\pi^2 \alphaem}{128\pi^2} 
    \int\!\frac{\CLrd l^2}{l^2} \int\!\frac{\CLrd m^2}{m^2} 
    D_{4,0}^{(1,+,+)}(\bm l, -\bm l,\bm m,-\bm m)
    \mathcal{F}_\mathrm{G}(\CLxP,l^2) 
    \mathcal{F}_\mathrm{G}(\CLxP,m^2)
\end{equation}
(we have suppressed the distinction between longitudinal and transverse
polarization). As shown in \cite{BW},
$D_{4,0}^{(1,+,+)}(\CLbfk_1,\CLbfk_2,\CLbfk_3,\CLbfk_4)$ can be written as a
sum of $D_{2,0}$ expressions with appropriate evaluation of momenta:
\begin{align} \label{D40contribs1}
  D_{4,0}^{(1,+,+)}(\CLbfk_1,\CLbfk_2,\CLbfk_3,\CLbfk_4)=g^2\frac{\sqrt2}{3}
  \bigl\{ \phantom{+} &
    D_{2,0}(\CLbfk_1,\CLbfk_2+\CLbfk_3+\CLbfk_4) +  
      D_{2,0}(\CLbfk_2,\CLbfk_1+\CLbfk_3+\CLbfk_4) + \notag\\
    +\, & D_{2,0}(\CLbfk_3,\CLbfk_1+\CLbfk_2+\CLbfk_4) + 
      D_{2,0}(\CLbfk_4,\CLbfk_1+\CLbfk_2+\CLbfk_3)- \notag\\
    -\, &  D_{2,0}(\CLbfk_1+\CLbfk_2,\CLbfk_3+\CLbfk_4) - 
      D_{2,0}(\CLbfk_1+\CLbfk_3,\CLbfk_2+\CLbfk_4)- \notag\\
    -\, & D_{2,0}(\CLbfk_1+\CLbfk_4,\CLbfk_2+\CLbfk_3)\,\bigr\} \, .
\end{align}
Here $\CLbfk_1,\dots,\CLbfk_4$ are the transverse components of the gluon
momenta at the lower end of the quark-loop. In our case the quark-loop is
coupled to the proton through the unintegrated gluon structure function
$\mathcal F_\mathrm G$ (cf.\,\eqref{DiffCrossSecT1} and
\eqref{DiffCrossSecL1}), and we have to substitute $\CLbfk_1=-\CLbfk_2=\CLbfl$
and $\CLbfk_3=-\CLbfk_4=\CLbfl'$ and integrate over $\CLbfl$ and $\CLbfl'$. To
define leading-twist and twist-four corrections we simply expand in powers of
$1/Q^2$.

Before we describe the formal expansion in inverse powers of $Q^2$, let us give
a qualitative description \cite{BW2}. We begin with the $p_t^2$-integral of the
transverse cross section \eqref{DiffCrossSecT1} and keep $M^2$ fixed.  As long
as $p_t^2$ is not small, the leading contribution of the $\bm l$-integral comes
from the region of small $l^2<p_t^2(M^2+Q^2)/M^2$ where $\mathcal{I}^{\,t}/l^2$
behaves as a constant. The $\bm l$-integral, therefore, simply leads to $\CLxP
g(\CLxP ,p_t^2(Q^2+M^2)/M^2)$, and the transverse cross section falls as
$[1/p_t^2\,\CLxP g(\CLxP,p_t^2(Q^2+M^2)/M^2]^2$. Using naive dimensional
arguments it follows that this region belongs to higher-twist.  Taking the
integral over $p_t^2$ one finds dominance of the low $p_t^2$ region where
perturbation theory breaks down. The leading-twist of the transverse cross
section is obtained by extending the convergent $p_t^2$-integral up to
infinity. The twist-four term arises as a correction: the correct upper limit
to the $p_t^2$-integral is $Q^2 (1-\beta)/ \beta$, i.e.\ we have to subtract
the integral from $Q^2 (1-\beta)/ \beta$ to infinity. This is the negative
twist-four correction to the transverse cross section. The convergence of the
$p_t^2$-integration implies also that there is no $\ln Q^2$ in the twist-four
correction to the transverse cross section.  Turning to the longitudinal cross
section, one notices in \eqref{DiffCrossSecL1} the extra $p_t^2$ factor in
front of the square brackets: this changes, compared to the transverse case,
the $p_t^2$-behavior of the cross section in two ways. When integrating over
$p_t^2$, the small $p_t^2$ region no longer dominates, whereas in the large
$p_t^2$-region we encounter a logarithmic divergence. This explains the
$1/Q^2$-suppression of the cross section and the appearance of a $\ln Q^2$.
The $\beta$ (or $M^2$) dependence of the two cross sections arises from a
closer inspection of the two formulae \eqref{DiffCrossSecT1} and
\eqref{DiffCrossSecL1}: whereas the transverse cross section (both leading- and
higher-twist) vanishes near $\beta=1$, the longitudinal one stays finite. This
leads to the conclusion that the longitudinal cross section (which for
$\beta\ll1$ is much smaller than the transverse one) dominates in the large
$\beta$-region. In other words, the diffractive $q\bar{q}$ cross section near
$\beta=1$ is mainly longitudinal and belongs to twist-four. Results of this
discussion are summarized in table~\ref{SignAndLogsToDiff1}.

In order to obtain quantitative expressions for the higher-twist cross sections
(and to verify our intuitive arguments), we return to the $p_t^2$ and
$\beta$-integrated expressions in \eqref{D40contribs1} and perform a formal
expansion in powers of $1/Q^2$. Using the ($\omega,\nu$)-representation for
$D_{2,0}$ in \eqref{D40contribs1} and expanding about the twist-four point
$\nu=-2$ yields for the transverse case
\begin{equation} \label{qqbarT}
  \frac1\omega D_{4,0}^{t(1,+,+)}(\omega,l^2,{l'}^2)^{\tau=4} =
  a_2^t\frac{4\pi\sqrt2}{3}\frac1\omega
  \sum_f e_f^2\frac{\sqrt8}{2\pi}\alphaS^2
  \left\{ 2l^4 + 2{l'}^4 - (\CLbfl+\CLbfl')^4 - (\CLbfl-\CLbfl')^4 \right\} \,.
\end{equation}
Similarly, in the longitudinal case we get
\begin{multline} \label{qqbarL}
  \frac1\omega D_{4,0}^{\ell(1,+,+)}(\omega,l^2,{l'}^2)^{\tau=4} =
  -b_2^\ell\frac{4\pi\sqrt2}{3}\frac1\omega
  \sum_f e_f^2\frac{\sqrt8}{2\pi}\alphaS^2
  \Bigl\{ 2l^4\ln l^2 + 2{l'}^4\ln {l'}^2 - \\ -
    (\CLbfl+\CLbfl')^4\ln(\CLbfl+\CLbfl')^2
    - (\CLbfl-\CLbfl')^4\ln(\CLbfl-\CLbfl')^2 \Bigr\} \,.
\end{multline}
Here, $l^2$ and ${l'}^2$ are normalized by $Q^2$.  Working in DLA, we have for
the unintegrated gluon structure function $\mathcal F_\mathrm G$ an expression
of the form
\begin{equation} \label{Fcal}
  \mathcal F_\mathrm G(x, l^2)=
   \frac{1}{Q^2} \frac{\partial}{\partial l^2} 
    \int\!\frac{\CLrd\omega}{2\pi\CLri} \left( \frac{1}{x} \right )^{\omega}
    \exp\left(\frac{\gamma_2}{\omega} \ln l^2/q_0^2\right)
    \,\varphi_2(\omega)^{\tau=2} \,,
\end{equation}
where $\varphi_2(\omega)$ is the initial distribution of the gluon structure
function and $q_0^2=Q_0^2/Q^2$. Multiplying \eqref{qqbarT} with functions
$\mathcal F_\mathrm G(l^2)$ and $\mathcal F_\mathrm G({l'}^2)$ and performing
the integrations over $\bm l$ and $\bm l'$, we arrive at the following
higher-twist correction to the differential diffractive cross section:
\begin{equation} \label{deltadiffcrosst}
  \left.\Delta \frac{\CLrd\sigma^{q\bar{q},t}}{\CLrd t}\right|_{t=0}= 
    -\frac{4\pi^2 \alphaem}{Q^4} \frac{a_2^t}{6\pi^2}
    \frac{\alphaS^2}\omega \sum_f e_f^2 
    \exp\left(2\frac{\gamma_2}{\omega} \ln Q^2/Q_0^2\right)
    [\varphi_2(Q_0^2,\omega)]^2 \,.
\end{equation}
For the longitudinal case the $\bm l$, $\bm l'$ integrals are slightly more
complicated, since \eqref{qqbarL} contains logarithms. In order to obtain the
maximum number of logarithms, we split the integration interval into
subintervals $l^2 < {l'}^2$ and ${l'}^2 < l^2$. The result is
\begin{equation}\label{deltadiffcrossl}
  \left.\Delta \frac{\CLrd\sigma^{q\bar{q},\ell}}{\CLrd t}\right|_{t=0}= 
    \frac{4\pi^2 \alphaem}{Q^4} \frac{b_2^{\ell}}{6\pi^2}
    \frac{\alphaS^2}{2\gamma_2} \sum_f e_f^2
    \left\{1-
      \exp\left(2\frac{\gamma_2}{\omega} \ln Q^2/Q_0^2\right)\right\}
    [\varphi_2(Q_0^2,\omega)]^2 \,.
\end{equation}

Before we can draw the relation with the inclusive structure functions we still
have to integrate over the momentum transfer $t$. Repeating the discussion in
the sequel of eqn.\,\eqref{initialasym}, we note that in DLA the nonzero
momentum transfer enters only into the lowest part of the ladders, the initial
distributions $\varphi_2$. Integration in \eqref{deltadiffcrosst} or
\eqref{deltadiffcrossl} over $t$, therefore, simply means replacing
$\varphi_2^2$ by $\int\CLrd^2\CLbfq\,\varphi_2^2$, i.e.\ by an effective new
initial condition. Moreover, if we allow for diffractive dissociation of the
target proton, this initial condition will again be modified into a new initial
condition which we denote by $\varphi_4^\mathrm{diff}(\omega)$:
\begin{equation} \label{deltasigmal}
  \Delta \sigma^{q\bar{q},\ell}=
    \frac{4\pi^2 \alphaem}{Q^2} \frac{b_2^\ell}{6\pi^2}
    \frac{\alphaS^2}{2\gamma_2} \sum_f e_f^2 \frac{Q_0^2}{Q^2}
    \left\{1-
      \exp\left(2\frac{\gamma_2}{\omega} \ln Q^2/Q_0^2\right)\right\}
    \varphi_4^\mathrm{diff}(Q_0^2,\omega)
\end{equation}
(and a similar expression for the transverse cross section). Further below we
shall discuss how this diffractive cross section contributes to the structure
functions $F_{\ell,t}$.  In particular, we will derive a relation between
$\varphi_4^\mathrm{diff}(\omega)$ and $\varphi_{4\CLrS,\CLrA}(\omega)$
introduced in \eqref{initialtotal}.
%
%
\begin{figure}
  \begin{center}
    \input{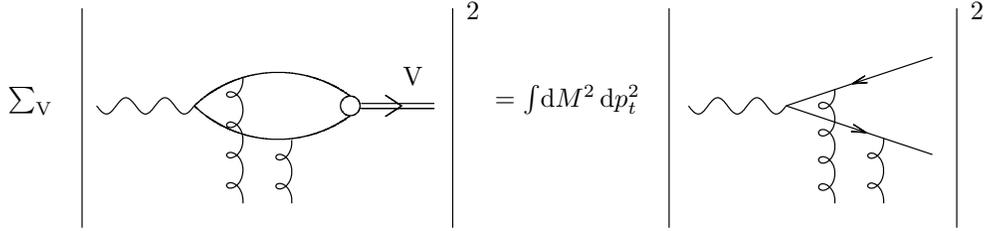}
    \caption{Duality relates the diffractive vector meson production cross
      section to the diffractive $q\bar q$-production cross section.}
    \label{Duality1}
  \end{center}
\end{figure}
\begin{figure}
  \begin{center}
    \input{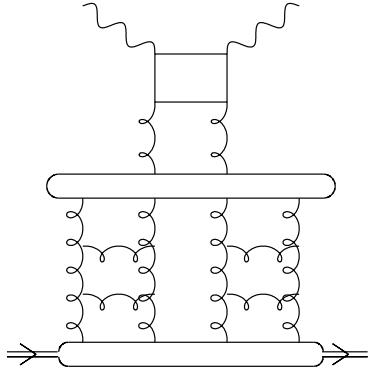}
    \caption{The (integrated) $q\bar qg$-production amplitude is given by
      an expression which is symbolized by this figure. Note that the effective
      $2\to4$ transition vertex is not the same as in fig.\,\ref{G_Lad1}b
      for $\DfourI$.
      }\label{qqgAmplitude1}
  \end{center}
\end{figure}
\begin{figure}
  \begin{center}
    \input{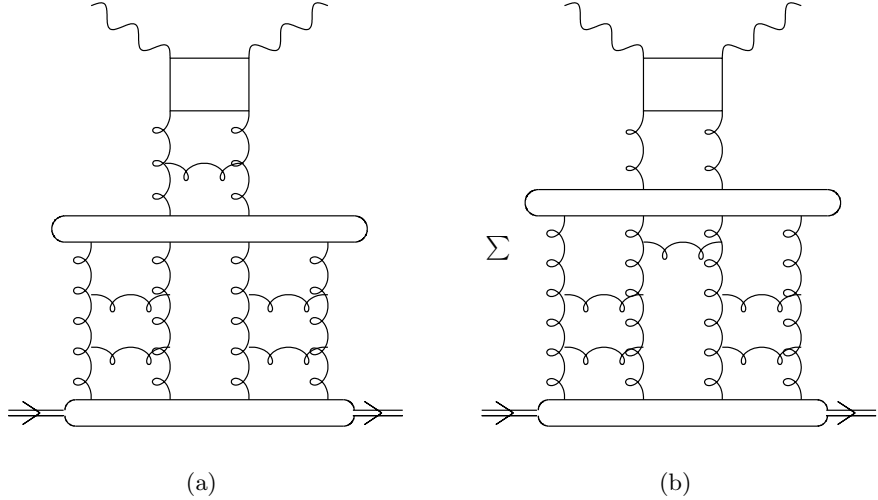}
    \caption{Same as in fig.\,\ref{qqgAmplitude1} but now for
      $q\bar qgg$-production. The sum denotes all possible couplings between
      systems (12) and (34).}\label{qqggAmplitude1}
  \end{center}
\end{figure}
%
%
\subsection{Diffractive Vector Meson Production}
Let us, now, turn to those diffractive processes which have been examined most
intensively both theoretically and experimentally, the diffractive production
of vector mesons: $\gamma^\ast +p\to V+p$, where $V$ can be any vector meson.
For the longitudinal photon this process has been shown to be calculable within
perturbative QCD \cite{R,BFGMS,CFS,MRT}:
\begin{equation} \label{DiffVec}
  \left. \frac{\CLrd\sigma_{\gamma^\ast p\to\mathrm Vp}^\ell}{\CLrd t}
    \right|_{t=0} =
  \frac{12\pi^3\alphaS^2}{N_\mathrm{c}^2\,\alphaem} \frac{M_\mathrm V}{Q^6}
  \Gamma_{\mathrm V\to e^+e^-}T(Q^2)\eta_\mathrm V^2
  \left|\left(1+\CLri\frac\pi2\frac\CLrd{\CLrd\ln x}\right)xg(x,Q^2)
  \right|^2 \,.
\end{equation}
Experiments show that the sum over all vector particles gives a significant
(about 20\,\%) contribution to the total diffractive cross section. As can be
seen in \eqref{DiffVec}, for an individual vector particle, the cross section
goes as $1/Q^6$. In ref. \cite{MRT} it has been shown that the production cross
section of a single vector particle can also be obtained if one starts with
open $q\bar{q}$-production and projects onto the corresponding angular momentum
and parity quantum numbers. This suggests to apply a simple duality argument
and to set, for the mass range $m_{\rho}<M<M_{c}$, the sum over vector particle
production cross sections \eqref{DiffVec} equal to the $M^2$- (and $p_t^2$-)
integrated cross section of open $q\bar{q}$-production \eqref{DiffCrossSecL1}
(we illustrate this equality in fig.\,\ref{Duality1}). In this way, the sum
over the vector particle cross sections turns into a twist-four contribution:
the additional factor $1/Q^2$ in \eqref{DiffVec} is due to the projection onto
the vector particle wave function.  As a result of this argument, we can use
the sum of the measured vector particle cross section to obtain a lower bound
of the integrated longitudinal $q\bar{q}$ cross section \cite{BS}, which has
been shown to belong to twist-four.
%
%
\begin{table}
  \begin{center}
    \begin{tabular}{l|ccc}
      Contribution & Sign & $\ln(Q^2/Q_0^2)$ & \\
      \hline
      Trans.\ $\tau=2$ & $+$ & $\ln(Q^2/Q_0^2)$ & non-pert.  \\
      Trans.\ $\tau=4$ & $+$ & $\ln(Q^2/Q_0^2)$ & \\
      Long.\ $\tau=2$ & $+$ & no & non-pert.\\
      Long.\ $\tau=4$ & $-$ & $(\ln(Q^2/Q_0^2))^2$ & 
    \end{tabular}
    \caption{The four contributions to diffractive $q\bar{q}g$-production.
    We list the sign structure and powers of $\ln Q^2$.}
    \label{SignAndLogsToDiff2}
  \end{center}
\end{table}
%
%
\subsection{Diffractive $q\bar{q}g$-Production and Multi-Jet Processes}
\begin{figure}
  \begin{center}
    \input{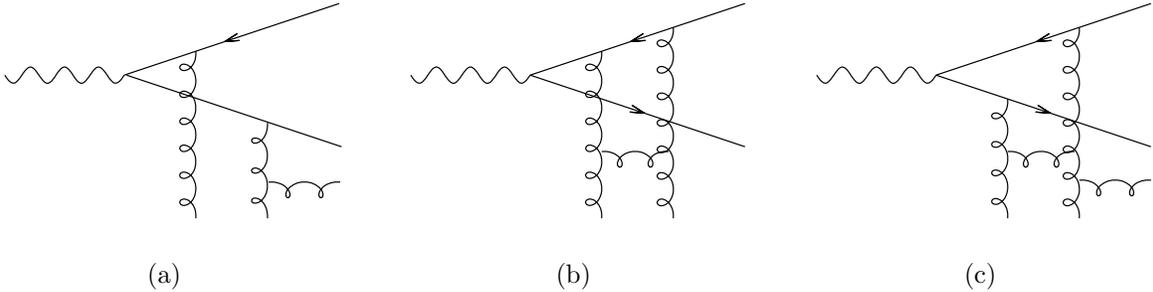}
    \caption{Three sample diagrams which contribute to $D_4$ at
      order $\alphaS^3$ ((a) and (b)) and at order $\alphaS^4$ (c). 
      Diagrams (a) and (c) enter also into diffractive
      $q\bar qg$-production at leading and next-to-leading order, resp.
      For the diffractive process the two $t$-channel gluons must certainly
      be in the color singlet, while there are also contributions from other
      color states which enter into $D_4$.} \label{qqgFeyn1}
  \end{center}
\end{figure}

Diffractive $q\bar{q}$-production is expected to be dominant near $\beta=1$.
If we go to $\beta <1$, we have to consider also gluon production processes
($q\bar qg$-, $q\bar qgg$-,$\dots$-jets). Cross sections for these processes
have been calculated in different kinematical regions: for very small $\beta$
(the triple Regge region) in \cite{BW,BJW}, for strong ordering in the
transverse momenta in \cite{Ryskin,LevinWuesthoff}.  The latter calculation
allows to describe the whole $\beta$-interval, but only the first one contains
both leading-twist and twist-four.  In this section, therefore, we make use
only of the results obtained in \cite{BJW} and discuss the higher-twist
contributions to diffractive $q\bar qg$-production. The calculations are done
in analogy to the $q\bar q$-production process discussed above: we keep all
powers in the transverse momenta, but we restrict ourselves to the leading log
in $M^2$ (or $1/\beta$).  An example of the Feynam diagrams is shown in
fig.\,\ref{qqgFeyn1}a. A complete description can be found in \cite{BJW}. In
general, the $s$-channel gluon can be emitted anywhere from one of the
$t$-channel gluons.

The analytic expressions for the cross section formulae \cite{BJW} are to
lengthy to be repeated here. We follow the same logic as in the sequel of
\eqref{DiffCrossSecT1} and \eqref{DiffCrossSecL1}. After integration over the
momenta of the outgoing quarks and the gluon the result \cite{BW} takes the
form illustrated in fig.\,\ref{qqgAmplitude1}: all diagrams can be drawn in
such a way, that at the upper end we have the fermion-loop $D_{2,0}$, and the
$s$-channel gluon is contained in an effective vertex function which is similar
(but not identical) to the $2 \to 4$ gluon vertex in \eqref{Vdef1}
(fig.\,\ref{G_Lad1}b). From \cite{BW} we take the expression:
\begin{align} \label{ThreeJetQuarkLoop1}
  \left.D_{q\bar qg,0}^{\ell,t}(\bm l,\bm l')\right|_{t=0}=\frac{3g^2}{4\sqrt2}
  D_{2,0}^{\ell,t}\otimes\bigl\{ &
    2G(\bm l,-\bm l)+2G(\bm l',-\bm l')+2G(\bm l,\bm l')+
    2G(\bm l,-\bm l')+\notag\\[-6pt] &
    +G(\bm l+\bm l',-\bm l-\bm l')+G(\bm l-\bm l',-\bm l +\bm l') -
    2G(\bm l,-\bm l+\bm l')\notag\\ &
    -2G(\bm l,-\bm l-\bm l')-2G(\bm l',\bm l-\bm l')-
    2G(\bm l',-\bm l-\bm l')
  \bigr\} \,.
\end{align}
Here $G(\bm a,\bm b)$ is the same function that enters into the convolution
$D_2\otimes V$ (eqns.\,\eqref{Bet_Salp1}, \eqref{Vdef1}). The ladders at the
lower end of this vertex, again, denote (leading-twist) gluon structure
functions \eqref{Fcal}.

Since we did not present the explicit formulae for the (unintegrated)
diffractive cross sections, we only briefly sketch the qualitative picture
(details can be found in \cite{BJW}). Let $\CLbfk_2$ be the transverse momentum
of the gluon, and $\CLbfk_1$, $-(\CLbfk_1+\CLbfk_2)$ those of the quarks. The
easiest way is to start with the hard region (i.e.\ the transverse momenta of
quarks and gluons are large), and to consider configurations where the momenta
are ordered: $k_2^2< k_1^2$. In this region, the transverse cross sections goes
as $\CLrd\sigma^t\sim1/[k_1^2k_2^4]\cdot[\CLxP g(\CLxP,k_2^2)]^2$ and the
longitudinal cross section behaves as
$\CLrd\sigma^\ell\sim1/[k_1^4k_2^4]\cdot[\CLxP g(\CLxP,k_2^2)]^2$.  By
dimensional counting, this region belongs to twist-four. Leading-twist can be
obtained if one realizes that the integration over the gluon momentum
$\CLbfk_2$ is dominated by the small momentum region, i.e.\ by nonperturbative
physics. The remaining $\CLbfk_1$-integral provides, for the transverse photon,
a $\ln Q^2$-enhancement, whereas for the longitudinal photon it does not
(table~\ref{SignAndLogsToDiff2}).

The derivation of the twist-four contributions, now, follows the same way as
for $q\bar{q}$-production. We use the $(\omega,\nu)$-representation for
\eqref{ThreeJetQuarkLoop1}, and the twist-four part (near $\nu=-2$) takes, for
the transverse photon, the form:
\begin{multline}
  \frac1\omega D_{q\bar qg,0}^t(\omega,l^2,{l'}^2)^{\tau=4} =
  -a_2^t\frac{9}{\pi}\frac1\omega
  \sum_f e_f^2\alphaS^3\frac{Q^2}{Q_0^2}
  \Bigl\{ 2l^4\ln l^2 + 2{l'}^4\ln {l'}^2 - \\ -
    \left[2l^4+2{l'}^4-8l^2{l'}^2\right]
    \ln\left[\max(l^2,{l'}^2)\right]
  \Bigr\} \,.
\end{multline}
Similarly, for the longitudinal photon we obtain:
\begin{multline}
  \frac1\omega D_{q\bar qg,0}^\ell(\omega,l^2,{l'}^2)^{\tau=4} =
  \frac{b_2^\ell}{2}\frac{9}{\pi}\frac1\omega
  \sum_f e_f^2\alphaS^3\frac{Q^2}{Q_0^2}
  \Bigl\{ 2l^4\ln^2 l^2 + 2{l'}^4\ln^2 {l'}^2 \\ -
    \left[2l^4+2{l'}^4-8l^2{l'}^2\right]
    \ln^2\left[\max(l^2,{l'}^2)\right]
  \Bigr\} \,.
\end{multline}
The integration over $\CLbfl$, $\CLbfl'$ is done in the same way as described
after \eqref{Fcal}. With \eqref{Fcal} for the unintegrated structure function
we find
\begin{equation}
  \left.\Delta\frac{\CLrd\sigma^{q\bar qg,t}}{\CLrd t}\right|_{t=0}=
  -\frac{4\pi^2\alphaem}{Q^4} \frac{9a_2^t}{16\pi^3}
  \frac{\alphaS^3}{2\gamma_2} \frac 1\omega \sum_f e_f^2
  \left\{ 1-
    \exp\left( \frac{2\gamma_2}{\omega}\ln\frac{Q^2}{Q_0^2} \right)
  \right\} \varphi_2^2(\omega)
\end{equation}
and
\begin{equation}
  \left.\Delta\frac{\CLrd\sigma^{q\bar qg,\ell}}{\CLrd t}\right|_{t=0}=
  \frac{4\pi^2\alphaem}{Q^4} \frac{9b_2^\ell}{32\pi^3}
  \frac{\alphaS^3}{2\gamma_2^2} \sum_f e_f^2
  \left\{ 
    \exp\left( \frac{2\gamma_2}{\omega}\ln\frac{Q^2}{Q_0^2} \right)
    -1 -\frac{2\gamma_2}{\omega}\ln\frac{Q^2}{Q_0^2}
  \right\} \varphi_2^2(\omega) 
\end{equation}
for the transverse and for the longitudinal photon, resp. Finally, performing
the integration over $t$ and replacing $\varphi_2^2$ by $Q_0^2
\varphi_4^\mathrm{diff}$ leads to our final expressions for the twist-four
corrections to the diffractive structure functions $\Delta
\sigma^{q\bar{q}g,t}$ and $\Delta \sigma^{q\bar{q}g,\ell}$.

An important feature of these results is the sign structure (summarized in
table~\ref{SignAndLogsToDiff2}). For both the transverse and the longitudinal
photon we find the opposite signs compared to $q\bar{q}$-production
(table~\ref{SignAndLogsToDiff1}). This implies that when we are calculating the
total twist-four corrections to the diffractive cross section -- either
transverse or longitudinal -- we are adding two terms ($q\bar{q}$ and
$q\bar{q}g$) of opposite signs. Moreover, the higher-twist corrections for the
transverse and for the longitudinal diffractive cross sections have opposite
signs. This clearly allows for the possibility of substantial cancellations, in
particular in the corrections to $F_2^\mathrm D=F_{\ell}^\mathrm D +
F_{t}^\mathrm D$. The longitudinal diffractive cross section may, therefore, be
a cleaner place to look for twist-four corrections.

It is not difficult to generalize this discussion to the production of more
gluons in the region of very small $\beta$. For example, for the (integrated)
production cross section of two gluons we have to calculate the diagrams shown
in fig.\,\ref{qqggAmplitude1}. As before, use has been made of
eqns.\,\eqref{D4R} and \eqref{D3Rel1}. In this way all diagrams in
fig.\,\ref{qqggAmplitude1} can be rearranged in terms of the vertex
\eqref{ThreeJetQuarkLoop1}, and they can be grouped into two classes.
Beginning with the fermion-loop at the top of the diagram, we have, in the case
of fig.\,\ref{qqggAmplitude1}a, one BFKL rung, then the vertex
\eqref{ThreeJetQuarkLoop1}, and further below the two-gluon ladders for the
unintegrated gluon structure functions. In case of fig.\,\ref{qqggAmplitude1}b,
there is no rung between the fermion-loop and the vertex
\eqref{ThreeJetQuarkLoop1}, but instead four possible rungs below the vertex at
the top of the gluon ladders. Similarly, for three gluons we have three
possibilities: two gluon rungs above the vertex, one rung above and one set of
rungs below, or two sets of rungs below.
%
%
\subsection{Connection with the Structure Functions $F_{\ell,t}$ and
the AGK Cutting Rules} \label{SecAGK}
After we have discussed the different diffractive cross sections, we, now, turn
to the question of how these contributions enter into the inclusive cross
section \eqref{D4crosssection}. An important ingredient are the AGK cutting
rules \cite{AGK}.

We first return to the total cross section \eqref{D4crosssection} and
recapitulate how the AGK rules work. Let us, for the moment, replace the proton
by a virtual photon, i.e.\ we consider the elastic scattering of a virtual
photon on another virtual photon. This process can be treated perturbatively.
For the time being we do not expand in powers of $Q_0^2/Q^2$, but consider the
Regge limit at large but finite $Q^2$.  In order to study the AGK rules, we
start with the discontinuity across the four-gluon $t$-channel intermediate
state. Disregarding all unnecessary details, this discontinuity takes the form
\begin{equation} \label{4gluondisc} 
  \CLdisc_{\omega}  F_{\gamma^* \gamma^*} \sim \frac{1}{4!} 
  C_4^* \otimes C_4
\end{equation}
where $F_{\gamma^* \gamma^*}$ denotes the $t$-channel partial wave for the
$\gamma^* \gamma^*$ scattering process, $C_4(k_1,k_2,k_3,k_4;\omega)=(\omega -
\sum \beta(k_i)) D_4(k_1,k_2,k_3,k_4;\omega)$ is the amputated
$\gamma^*\gamma^*\to\,$4-gluons partial wave, and the symbol $\otimes$ contains
all phase space factors. For $C_4$ we have the same decomposition
(eqn.\,\eqref{D4sumD4R_D4I}) as for $D_4$. In particular, $C_4^\mathrm I$ is
completely symmetric, and $C_4^\mathrm R$ contains a symmetric piece and a
piece with mixed symmetry (cf.\ the discussion after \eqref{D4R}). In the
unitarity integral \eqref{4gluondisc}, we take from both $C_4$ and $C_4^*$
either the symmetric pieces or the pieces with mixed symmetry, i.e.\ there is
no interference term.

First consider the symmetric pieces, e.g.\ $C_4^\mathrm I$ for both factors in
\eqref{4gluondisc}. Writing $C_4^\mathrm I$ in terms of the auxiliary
potentials $T_{(ij)(kl)}$ from our discussion above, we have in our unitarity
integral on the r.h.s.\ of \eqref{4gluondisc} the three identical `diagonal'
terms $T_{(12)(34)}\otimes T_{(12)(34)}$, $T_{(13)(24)}\otimes T_{(13)(24)}$,
and $T_{(14)(23)}\otimes T_{(14)(23)}$.  Together with the statistical factor
$1/4!$, this gives a weight factor $1/2(1/2)^2$ for the diagonal term (this
counting does not include color). Similarly, `nondiagonal' terms,
$T_{(12)(34)}\otimes T_{(13)(24)}$ etc., yield the weight factor $(1/2)^2$.  In
order to verify the AGK cutting rules we compare $2\,\mathrm{Im}\, T_{\gamma^*
  \gamma^*}$ of this amplitude with the different cuts, $\sigma_0$, $\sigma_1$,
and $\sigma_2$ (the subscripts refer to the number of cut ladders).  The
(diagonal) ``diffractive cut'' $\sigma_0$ has the cutting line between the
gluon lines `2' and `3'. Its contribution is positive and has the weight
$(1/2)^2$ (due to the statistical factors inside the ladders). For the ``double
multiperipheral'' cut $\sigma_2$ the cutting line runs also between gluons `2'
and `3', but its (positive) contribution has the weight $2(1/2)^2$. Finally,
the ``absorptive cut'' $\sigma_1$ is negative; the cutting line runs between
line `1' and `2' or `3' and `4', and the counting gives $-2\cdot3\cdot1/3!=-1$.
Adding all these contributions, we arrive at $-(1/2)^2$, i.e.\ minus the
diffractive cross section $\sigma_1$. The result agrees with $2\,\mathrm{Im}\,
T_{\gamma^* \gamma^*}$, as anticipated by the AGK rules.  Two features of these
results are important for us: in total, the four-gluon $t$-channel state gives
a {\it negative} correction to the single-ladder contribution (this justifies
the minus sign in \eqref{D4crosssection}), and its absolute value equals the
`diffractive cut' $\sigma_0$. It is easy to verify that the same conclusion
also holds for nondiagonal pieces $T_{(12)(34)}\otimes T_{(13)(24)}$ etc.

Next, the part with mixed symmetry: in \eqref{4gluondisc} we put $C_4^\mathrm
R$ for both $C_4$-factors, and we take the terms with the anti-symmetric
structure constants ($f^{abc'}f^{c'cd}$ etc.). Their color and momentum
structure is the same as in \eqref{initialasym}. They are connected with the
odd signature reggeized gluon.  Again, in the unitarity integral we have
diagonal terms of the type $f_\CLrA(12,34)\otimes f_\CLrA(12,34)$. Their total
weight is $2(1/2)^2$. In the counting of the nondiagonal terms we encounter
several cancellations: the weight factor turns out to be $1/2$.  Now, we
compare this with the cuts $\sigma_0$, $\sigma_1$, and $\sigma_2$. For the
diagonal $\sigma_0$ and $\sigma_2$ contributions where the cutting line runs
between gluon `2' and `3', we find the weight factors $(1/2)^2$ and $2(1/2)^2$,
resp. They are the same as the counting factors for the even signature ladders
which we have discussed before.  In particular, we again have the negative sign
for the sum of the three cuts. There is, however, an important difference
between even and odd signature reggeons: for odd signature the real part of the
signature dominates, and the leading contribution $\sigma_0$ is contained in
the LO BFKL equation.  The only contribution to $\sigma_0$ from the four-gluon
intermediate state comes from the `diagonal' pieces which present the two gluon
contributions to the gluon trajectory functions. Consistency, therefore,
requires that the magnitude of this term is linked to the leading order BFKL
ladder. One can check from \eqref{D4R} that this condition is, indeed,
satisfied.

In summary, the AGK rules allow us to determine the sign of our initial
conditions. We return to deep inelastic scattering off the proton, consider the
twist-four term in the expansion in powers of $Q_0^2/Q^2$ and replace, in
\eqref{4gluondisc}, the second $C_4$ factor by our general ansatz
\eqref{initialtotal}, i.e.\ by our nonperturbative initial conditions
$\varphi_{4\CLrS}$ and $\varphi_{4\CLrA}$. In order to preserve the sign
structure dictated by the AGK rules, both functions have to be positive.
Moreover, since $\varphi_{4\CLrA}$ is connected with the reggeization of the
gluon trajectory function, it is linked to the size of the twist-four part in
the BFKL ladder, $\Delta F^{(2)}$. In this first attempt to estimate the
higher-twist contribution, we will consider two different values for
$\varphi_{4\CLrA}$ which we expect to present a reasonable range.

Next we attempt to relate these unknown initial conditions to the diffractive
$q\bar{q}$ cross section \eqref{deltasigmal}, in order to obtain an estimate of
their magnitude.  The easiest way is a comparison of the two-ladder diagrams in
$\DelFfour$ with the diffractive cross section (fig.\,\ref{Diff_Lad1}).  We use
the coupling \eqref{initialtotal}, take the color singlet states of the two
ladders, project onto the twist-four term and compute the ``diffractive'' cut
$\sigma_0$ between the gluon lines 2 and 3.  Equating the result with the cross
section \ref{deltasigmal} we arrive at the equation:
\begin{equation} \label{RelPhiDiff_PhiSA1}
  \frac{5}{4} \varphi_{4\CLrS} - \frac{3}{4} \varphi_{4\CLrA} =
  \varphi_4^\mathrm{diff}.
\end{equation}

This result is quite remarkable, since -- at first sight -- it does not seem to
agree with the AGK rules. Namely, naively one might have expected that
$\varphi_{4\CLrS}=\varphi_4^\mathrm{diff}$, i.e.\ the four-gluon contribution
equals (up to the overall minus sign which we have extracted already in
\eqref{D4crosssection}) the diffractive cross section.  Let us recapitulate the
origin of the l.h.s.\ of \eqref{RelPhiDiff_PhiSA1}: from \eqref{D4sumD4R_D4I}
we deduced that QCD diagrams for $T_{\gamma^* \gamma^*}$ with four gluons in
the $t$-channel have to be decomposed according to their symmetry under
permutation of color and momenta (the antisymmetric terms have their origin in
the reggeization of the gluon). The same argument forces us to use the ansatz
\eqref{initialtotal}-\eqref{initialasym}, i.e.\ we have the two functions
$\varphi_{4\CLrS}$ and $\varphi_{4\CLrA}$. The AGK rules work for each term
separately, but the $t$-channel state with four reggeized gluons couples only
to $\varphi_{4\CLrS}$. Finally, when computing the two-ladder diagrams in
$T_{\gamma^* \gamma^*}$, we find for the coupling of two ladders to the proton
the combination on the l.h.s.\ of \eqref{RelPhiDiff_PhiSA1} (the fact that the
factor in front of $\varphi_{4\CLrS}$ equals $5/4$ and not $1$ is a result of
the presence of the second and third term in \eqref{initialsym}).

There are two important consequences of \eqref{RelPhiDiff_PhiSA1}. First, from
this single condition we cannot fix both initial conditions $\varphi_{4\CLrS}$
and $\varphi_{4\CLrA}$. Secondly, due to the minus sign on the l.h.s.,
$\varphi_{4\CLrS}$ can be larger than the ``naive'' expectation
$\varphi_{4\CLrS}=\varphi_4^\mathrm{diff}$. We propose to proceed as follows.
The longitudinal cross section for diffractive $q\bar{q}$-production allows to
estimate $\varphi_4^\mathrm{diff}$. From our discussion above we know that
$\varphi_{4\CLrA}$ is connected with the reggeization of the gluon: in our
numerical analysis we will vary $\varphi_{4\CLrA}$ in such a way that its
correction to the leading-twist structure function ranges between $20\,\%$ and
$100\,\%$ (see below). Together with the variation of $\varphi_{4\CLrA}$ we
also find a change in the strength of $\varphi_{4\CLrS}$ and $\DelFI$, the most
interesting higher-twist contribution in the small $x$ region.
%
%
\subsection{The Sign Structure}
It may be helpful to briefly recapitulate the sequence of our arguments, in
particular the sign structure. The easiest way is to start with the twist-four
corrections to the diffractive cross sections; their signs are summarized in
tables~\ref{SignAndLogsToDiff1} and \ref{SignAndLogsToDiff2}.  Starting with
$q\bar{q}$-production, in the transverse case twist-four is a correction to
leading-twist and has a negative sign. The longitudinal cross section starts
with twist-four and is positive. Next diffractive $q\bar{q}g$-production: all
signs change, compared to $q\bar{q}$-production. This is due to the vertex
\eqref{ThreeJetQuarkLoop1} which has an intrinsic minus sign.

In the next step we have argued that when generalizing to an arbitrary number
of gluons in the diffractive final state and allowing for evolution in the
unintegrated gluon structure function we generate all diagrams contained in
$\DelFfour$: although there is no direct correspondence between the splitting
$\DelFfour=\DelFR+\DelFI$ and the different diffractive final states. As a memo
we nevertheless can say that $\DelFR$ ``is related to'' (i.e.\ gets the same
sign as) the cross section for $q\bar{q}$-production, and a similar
correspondence holds for $\DelFI$ and diffractive $q\bar{q}g$-,
$q\bar{q}gg$-$,\dots,$-production.

In the last step we have to remember that it is not only the diffractive states
which contribute to the structure functions. Making use of the AGK rules we
conclude that taking into account all other energy discontinuities simply means
changing the sign of the (total) diffractive contributions. The results are
summarized in table~\ref{TabSignF2}.
%
%
\begin{figure}
  \begin{center}
    \input{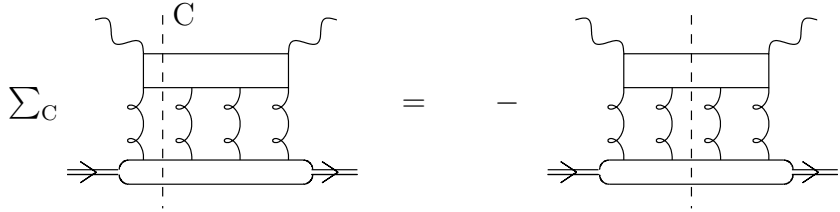}
    \caption{The AGK rules: The sum over the three cuts, which is necessary
      for the computation of the inclusive structure function, gives a
      contribution that is equal to minus one times the term where only
      the diffractive cut is applied.}\label{FigAGK1}
  \end{center}
\end{figure}
%
%
\section{Numerical Results} \label{SecNumRes}
\subsection{General Procedure and Determination of Initial Parton 
Distributions} \label{SecRelContribs}
Using the framework outlined above we have performed a numerical analysis. The
first step to be discussed is how we fix the free parameters in the various
initial distributions $\varphi$. From our discussion it should have become
clear that we have to determine four initial distributions, viz.\ $\varphi_2$
for $\DelFtwo$, $\varphi_3$ for the three-gluon correction to $\DelFtwo$, and
$\varphi_4$ which couples to both $\DelFR$ and $\DelFI$. Moreover, $\varphi_4$
consists of a symmetric part and a part with mixed symmetry, cf.\ 
eqn.\,\eqref{initialtotal}. All initial distributions depend on $\omega$, and a
pole at some positive $\omega$-value corresponds to a rising (in $1/x$) initial
condition.

Let us begin with a detailed discussion of the four-gluon case. As indicated
before, we relate $\varphi_4$ directly to the longitudinal diffractive
$q\bar{q}$ cross section (eqs.\,\eqref{deltasigmal} and
\eqref{RelPhiDiff_PhiSA1}), and the latter can be estimated with the help of
the measured cross section of longitudinal vector production if we sum over all
vector mesons in the final state \cite{BS}. Let us assume that -- after
integration over $t$, $p_t^2$ and $M^2$, and including diffractive dissociation
of the proton -- this cross section reaches a few per cent of the inclusive
(transverse) leading-twist cross section at $x=10^{-3}$, $Q^2=10\,$GeV$^2$.
Using this estimate we fix the initial condition $\varphi_4^\mathrm{diff}$ and,
with the help of \eqref{RelPhiDiff_PhiSA1}, also the combination of the initial
conditions $\varphi_{4\mathrm S}$ and $\varphi_{4\mathrm A}$, which enter into
the four-gluon amplitude. Since this procedure fixes only the difference
$5\varphi_{4\mathrm S}-3\varphi_{4\mathrm A}$ (cf.\,\eqref{RelPhiDiff_PhiSA1})
we still have freedom for determining the ratio of $\varphi_{4\mathrm S}$ and
$\varphi_{4\mathrm A}$.  In a perturbative toy model, namely deep inelastic
scattering off a virtual photon, $\varphi_{4\mathrm A}$ is linked to the
leading-twist amplitude: it provides the $t$-channel four-gluon state inside
the BFKL amplitude (cf.\ the discussion in section~\ref{SecAGK}) and, hence,
can be derived from the size of the leading-twist contribution.  However, when
replacing the (target) virtual photon by the proton, this connection becomes
somewhat uncertain, and we, therefore, consider different values for the ratio
$\phifA/\phifS$.  Starting with the toy model, we find it reasonable to expect
that $\varphi_{4\mathrm A}$ should not be smaller than $\varphi_{4\mathrm S}$.
We, therefore, set $\phifA=\lambda\phifS$ and consider the range
$1.0<\lambda<1.5$.  Eqn.\,\eqref{RelPhiDiff_PhiSA1} indicates that this
seemingly small variation will nevertheless have a large effect: if $\lambda$
reaches the vicinity of $5/3\approx1.6$, $\varphi_{4\mathrm S}$ increases
strongly (for fixed $\varphi_4^\mathrm{diff}$).  This enhances the contribution
of $\DelFI$, whereas $\DelFR$ changes much less, since it depends on both,
$\varphi_{4\mathrm S}$ and $\varphi_{4\mathrm A}$ (cf.\,\eqref{deltaFR}). Since
$\DelFI$ and $\DelFR$ enter with different signs, a variation of $\lambda$
between $1$ and $1.5$ will have a large effect on the total twist-four
contribution.

Having fixed the initial conditions $\varphi_4$ we can, in principle, predict
other higher-twist corrections to the diffractive cross section. In a future
analysis of diffractive jet data at HERA it may be possible to measure the
twist-four $q\bar{q}$ and $q\bar{q}g$ contributions directly; this will provide
an important consistency check of this strategy.  As we have said before,
because of the cancellations between the $q\bar{q}$ and $q\bar{q}g$ parts the
total higher-twist contribution to $F_2^\mathrm D$ may be smaller than the
naive estimate based on longitudinal $q\bar{q}$ production.

As to the other two initial conditions, $\varphi_2$ and $\varphi_3$, we have
not been able to relate them to any direct measurement. Starting with
$\varphi_2$, for simplicity we assume that at the scale $Q_0^2$ the initial
distribution for twist-four is the same as for the leading-twist case. It
follows that the transverse contribution is positive, whereas the longitudinal
twist-four part of the two-gluon amplitude contributes with a negative sign to
$F_2$, since $b_2^\ell$ in eqn.\,\eqref{D2long1} is negative. For the
three-gluon amplitude we make an even more ad hoc choice and demand that at
$x=10^{-3}$, $Q^2=1\,$GeV$^2$ its absolute magnitude in $\Delta F_\ell$ equals
the mean value of $\DelFtwo$ and $\DelFR$. Our numerical analysis shows that
the choice of the point, $x=10^{-3}$, $Q^2=1\,$GeV$^2$, hardly influences our
results, since $\DelFtwo$, $\DelFthree$ and $\DelFR$ have nearly the same shape
(up to factors $\sqrt{\alphaS(Q^2)}$ and $\alphaS(Q^2)$). In principle, for
both, $\varphi_2^{\tau=4}$ and the three-gluon term $\varphi_3$, not only the
absolute magnitude but also the overall signs are undetermined. We have fixed
these signs with the help of the following considerations. First, in $\DelFtwo$
there is no reason why there should be a sign change when going from
leading-twist to twist-four. As an example, in our toy model described above,
deep inelastic scattering of two virtual photons with virtualities $Q_1^2 \ll
Q_2^2$, both leading- and higher-twist have the same sign. Next, as we have
argued above, $\DelFthree$ and $\DelFR$ are supposed to be related to the NLO
and NNLO corrections to BFKL\@. Our choice of signs is consistent with the
expectation that these corrections have alternating signs, i.e.\ $\DelFthree$
is negative with respect to $\DelFtwo$, and $\DelFR$ is positive again. The
sign of $\DelFR$, on the other hand, has to be consistent with \eqref{deltaFR}
and the AGK cutting rules, i.e.\ posititive for the transverse part. All in all
we, therefore, believe that our choice of signs is on a rather safe ground.
Table~\ref{TabSignF2} summarizes the sign structure of the various leading- and
next-to-leading twist parts in $F_2$, cf.\,\eqref{StrucFuncDefEq1}.

As one of the results of our analysis we observe that $\DelFR$ and $\DelFI$
enter with opposite signs and, depending on the choice of parameters, tend to
weaken each other: although we have some freedom for determining the relative
contribution of $\varphi_{4\mathrm S}$ and $\varphi_{4\mathrm A}$, both,
$\DelFR$ and $\DelFI$, are sizeable and enter with opposite signs.  But despite
these cancellations the total twist-four contribution is not small at
$Q^2=1\,$GeV$^2$. It is negative, and the contribution of the transverse photon
is substantially larger than that of the longitudinal one.
\begin{table}
  \begin{center}
    \begin{tabular}{l|ccccc}
      & $F_2^{\tau=2}$ & $\DelFtwo$ & $\DelFthree$ &
        $\DelFR$ & $\DelFI$ \\[2pt]
      \hline
      long.  & $+$ & $-$ & $+$ & $-$ & $+$ \\
      trans. & $+$ & $+$ & $-$ & $+$ & $-$
    \end{tabular}
    \caption{The various contributions to $F_2$ enter with different signs.}
    \label{TabSignF2}
  \end{center}
\end{table}
%
%
\begin{figure}
  \begin{center}
    \input{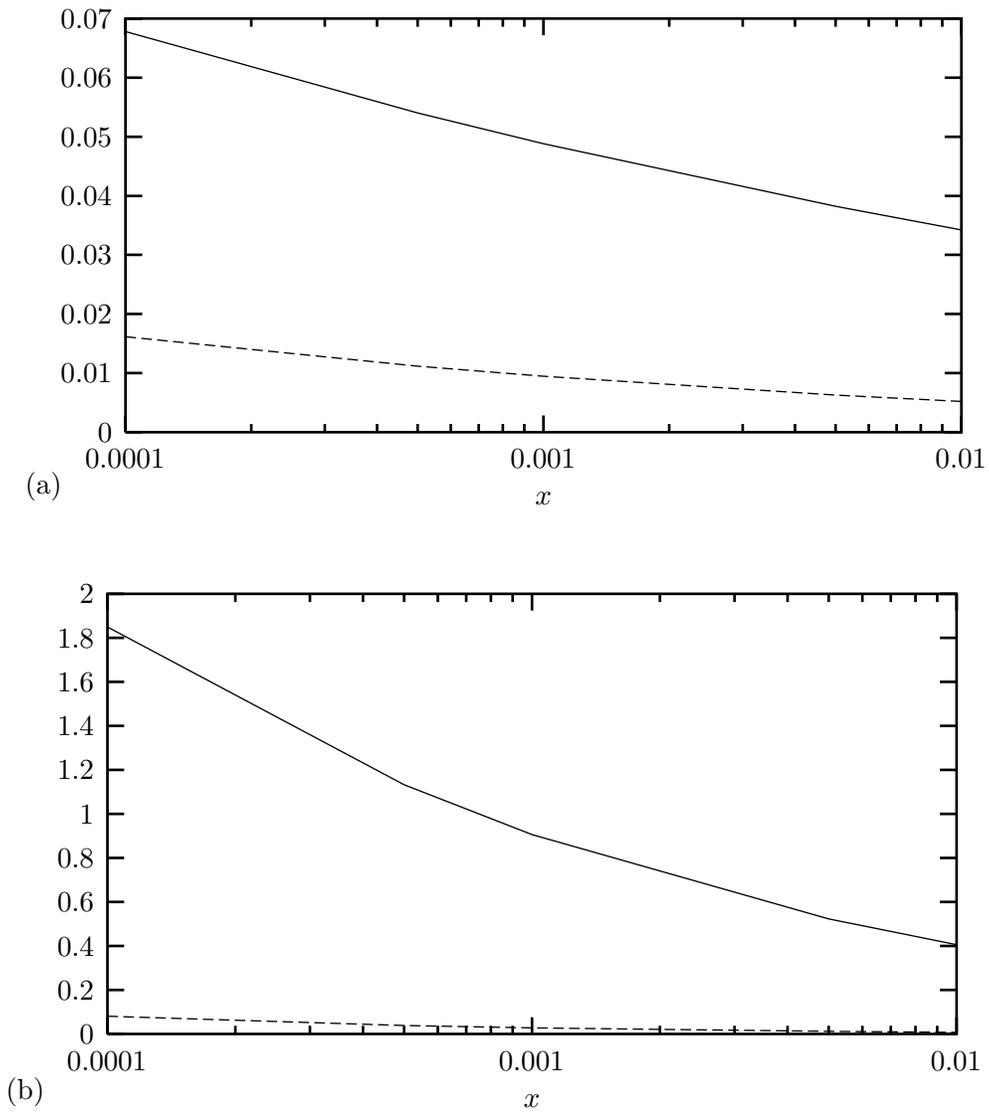}
    \caption{Transverse leading-twist structure function in DLA (solid line)
      in comparison with the longitudinal twist-four contribution to $F_2$ of
      the diffractive $q\bar q$-production process as a function of
      $x$ at (a) $Q^2=1\,$GeV$^2$ and (b) $Q^2=10\,$GeV$^2$. For this scenario,
      we have chosen constant initial distributions at $Q_0^2=0.5\,$GeV$^2$.
      } \label{Const_plot01ab}
  \end{center}
\end{figure}
%
%

In our numerical analysis we have used our formulae in order to perform, in the
kinematical regime of HERA, a ``semiquantitative'' numerical study of the
potential r\^ole of twist-four at small $x$ and low $Q^2$. As we have
emphasized several times the limitations due to the DLA are severe. On the
leading-twist level it is known that this approximation, which in the anomalous
dimension retains only the most singular (near $\omega=0$) term, is
insufficient to describe HERA data at small $x$ and low $Q^2$: it is, at least,
necessary to keep also the constant term within the gluon anomalous dimension.
We, therefore, expect that a similar level of accuracy will be needed also for
a realistic estimation of twist-four corrections.  In order to minimize the
inaccuracy we present the ratios of the twist-four contributions to the DLA
leading-twist transverse structure function.  In this way we hope that our
results describe the trend correctly on a semiquantitative level. For
leading-twist we are fixing the free parameter within the initial distribution
by demanding that at $x=10^{-3}$, $Q^2=10\,$GeV$^2$ our DLA expression for
$F_2$ is equal to the corresponding MRRS \cite{MRRS} value.

As outlined above, we are using the diffractive vector production cross section
to estimate twist-four contributions to $F_2$. To be precise, we assume that
the longitudinal twist-four $q\bar{q}$ cross section \eqref{deltasigmal}
reaches three per cent at $Q^2=10\,$GeV$^2$, $x=10^{-3}$ relative to the
transverse leading-twist contribution. As to the choice of the starting scale
$Q_0^2$, our limitations due to the DLA are causing a particular problem. In
DLA, the four-gluon amplitude $\DelFI$ vanishes at $Q^2=Q_0^2$, and it needs
some $Q^2$ evolution before it ``forgets'' about its starting condition. In an
analysis which goes beyond the DLA, we would start with some nonzero value. In
order to escape from this unrealistic behavior near $Q_0^2$, we have chosen a
very small value for the input scale, viz.\ $Q_0^2=0.5\,$GeV$^2$, and hope that
for $Q^2$ values greater than, say, $0.8\,$GeV$^2$, these unwanted effects are
suppressed. For the computation of the double Mellin transforms we have used
the method outlined in ref.\cite{Lyness1}, which proves to be very accurate and
efficient. The numerical results of our analysis also strongly depend upon the
$x$-shape of the initial distributions, both for the leading-twist and for the
twist-four correction. We, therefore, compare three different scenarios, two
with constant initial conditions and with different choices for the ratio
$\varphi_{4\mathrm A}/\varphi_{4\mathrm S}$, and one with rising initial
conditions.
%
%
\begin{figure}
  \begin{center}
    \input{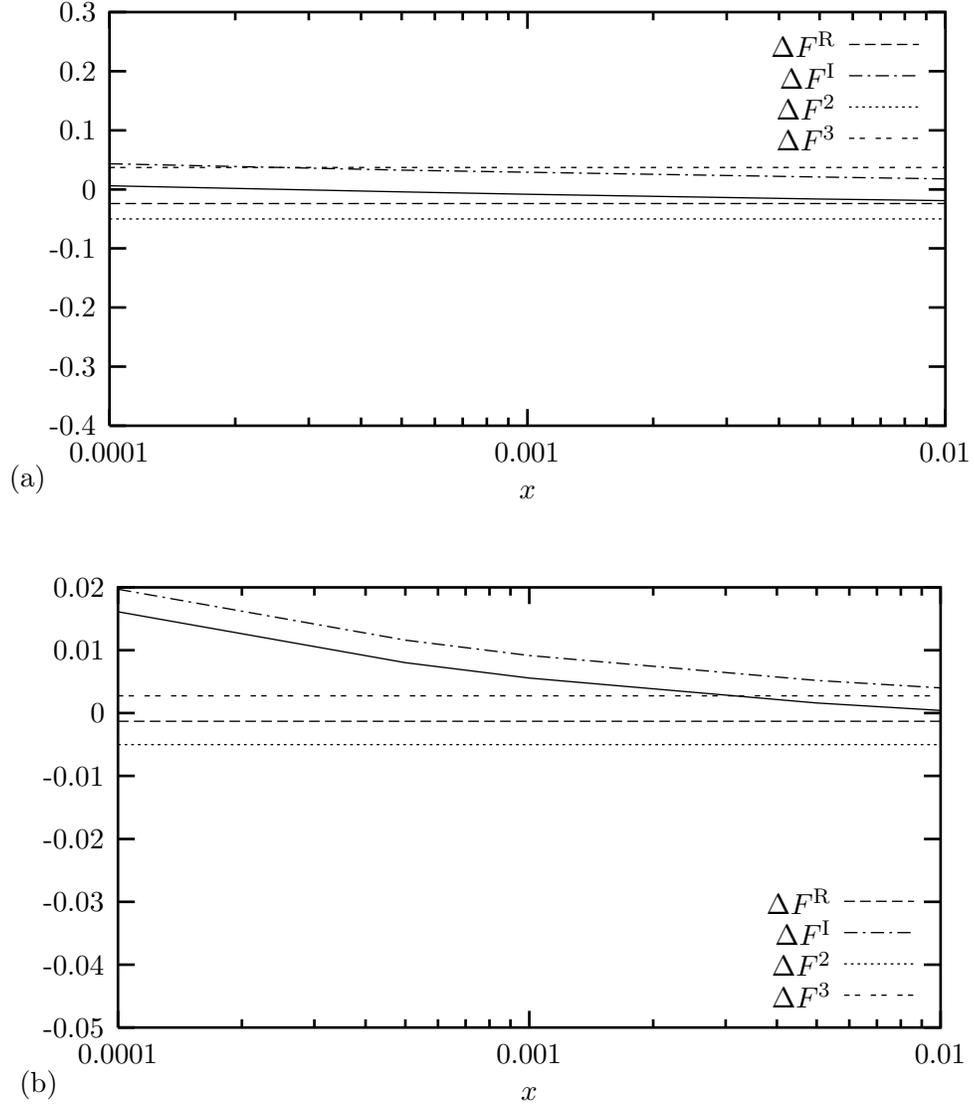}
    \caption{Longitudinal part of the twist-four contributions
      (solid line) and the four terms they are composed of
      as a function of $x$ at (a) $Q^2=1\,$GeV$^2$ and
      (b) $Q^2=10\,$GeV$^2$.
      Normalization with transverse leading-twist values.
      } \label{Const_plot02ab}
  \end{center}
\end{figure}
\begin{figure}
  \begin{center}
    \input{plot03.pst}
    \caption{Same as in fig.\,\ref{Const_plot02ab} but, now, for the
      transverse twist-four contributions.
      } \label{Const_plot03ab}
  \end{center}
\end{figure}
%
%
\subsection{Constant Initial Parton Distribution}
Following a presently popular trend in analysing HERA data, we first assume
that the leading-twist gluon structure function at the input scale $Q_0^2$ is
flat at small $x$. In two subsections we will consider the results for two
different choices of the ratio $\lambda\equiv\phifA/\phifS$, viz.\ $\lambda=1$
and $\lambda=1.5$. First, we show in figs.\,\ref{Const_plot01ab}a
and~\ref{Const_plot01ab}b the $x$-dependence of the transverse leading-twist
structure function in comparison with the twist-four contribution due to
diffractive longitudinal $q\bar{q}$-production. In the lower figure the point
$x=10^{-3}$ is our point of comparison: the higher-twist term has reached
$3\,\%$ of $F_2$.  The comparison of the two figures shows the expected
behavior: higher-twist becomes more important at low $Q^2$. If there would be
only this twist-four contribution, we would have a positive $20\,\%$ correction
at $Q^2=1\,$GeV$^2$.
%
%
\subsubsection{Weak Anti-Symmetric Initial Distribution $(\lambda=1)$}
The picture changes quite substantially if we include the other twist-four
contributions. We start with a discussion of the longitudinal part.
Fig.\,\ref{Const_plot02ab}a shows how this twist-four part is composed of the
four contributions (the twist-four part of the two-gluon amplitude $\DelFtwo$,
$\DelFthree$ and the four-gluon contributions, $\DelFR$ and $\DelFI$) at
$Q^2=1\,$GeV$^2$. In order to simplify comparisons, we have normalized all
twist-four contributions to the corresponding DLA transverse leading-twist
values. As can be seen, two of the four contributions ($\DelFthree$ and
$\DelFI$) enter with positive and the remaining two with negative signs, so
that the sum (solid line) gives a very small contribution. Let us take a closer
look at the cancellations, in particular those between $\DelFR$ and $\DelFI$.
As we argued before, after we have set $\phifA=\phifS$, the relative sign as
well as the relative magnitude are fixed: in fig.\,\ref{Const_plot02ab}a (at
$Q^2=1$\,GeV$^2$) $\DelFR$ and $\DelFI$ almost compensate each other. Due to
\eqref{D4_Sad_Approx1} $\DelFI$ will become dominating at very small $x$. In
our calculations this region is not yet reached.

The sign pattern of the individual contributions does not change when we go to
higher $Q^2$-values (fig.\,\ref{Const_plot02ab}b), although the total sum, now,
becomes positive in the full $x$-interval. A look at the absolute values, which
are not shown here, tells us that the twist-four contributions have actually
increased compared to the case at $Q^2=1\,$GeV$^2$ -- the $1/Q^2$ suppression
sets in rather late.  But since leading-twist has increased as well, the ratio
of the two decreases. Moreover, since $\DelFI$ is the twist-four contribution
with the strongest evolution (cf.\,\eqref{D4_Sad_Approx1}) the total twist-four
contribution is positive and reaches up to 1.5\,\% of leading-twist
(fig.\,\ref{Const_plot02ab}b). As to the other pair of contributions
($\DelFtwo$ and $\DelFthree$), they also come with opposite signs and,
approximately, the same strength. But this clearly depends very much on our
choice of the initial conditions. For example, a smaller choice of $\varphi_3$,
the initial distribution of the three-gluon amplitude, which, as outlined in
section~\ref{SecRelContribs}, cannot be related to any experimental
measurement, results in significant smaller values of $\Delta F_\ell$.

Next we take a look at the transverse contributions
(fig.\,\ref{Const_plot03ab}). As always in our discussion, this contribution
represents the DLA, and as we discussed before, it is suppressed by one power
of $\ln (Q^2/Q_0^2)$ in comparison to the longitudinal (twist-four) photon.
Nevertheless, our results for $\Delta F_t$ represent the leading contributions
for the transverse structure function. We plot our results in the same way as
we did for the longitudinal case, and the $x$ distributions for the two values
$Q^2=1\,$GeV$^2$ and $Q^2=10\,$GeV$^2$ are shown in
figs.\,\ref{Const_plot03ab}a and~b, resp. First, one notices that for each
separate twist-four contribution (again normalized to the leading-twist term)
the transition from $Q^2=1$\,GeV$^2$ to $Q^2=10$\,GeV$^2$ results in a much
stronger decrease than in the longitudinal case. This is certainly due to the
absence of the $\ln Q^2$-factor, which makes the $Q^2$-evolution weaker.
Altogether, at $Q^2=1\,$GeV$^2$ the transverse twist-four terms, which enter
into the correction to $F_t$ are of much larger size than in the longitudinal
case and the sum of all four terms, now, gives a negative contribution.

In our scenario the transverse twist-four contribution
(fig.\,\ref{Const_plot03ab}a) gives a negative correction which at $Q^2=1$
GeV$^2$ and $x=10^{-4}$ reaches 15\% of the leading-twist.  The main reason for
this is that $\DelFI$ is the dominating contribution.  If we would go to even
smaller values of $x$ the dominance of $\DelFI$ would become even more
pronounced but one has to keep in mind that in this case even higher-twist
contributions (twist-six, etc.) will, very likely, become important, as well.
At $Q^2=10\,$GeV$^2$ (fig.\,\ref{Const_plot03ab}b) the transverse twist-four
contribution has gone down to less than 4\%.  Compared to the longitudinal
case, the twist-four corrections of the transverse photon clearly dominates,
and we expect that this remains true even if we go beyond the DLA\@. We can use
our numerical results to determine the contibution of $\phifA$ within $\DelFR$:
using \eqref{deltaFR} and the numerical values in fig.\,\ref{Const_plot03ab} we
deduce that (for the transverse photon at $Q^2=1\,$GeV$^2$) it reaches about
20\,\% of leading-twist. We believe that this is a rather low value.  In order
to further decrease $\phifA$, we would have to choose $\lambda<1$. This seems
to be rather unlikely if we can believe the results of our toy model.
\begin{figure}
  \begin{center}
    \input{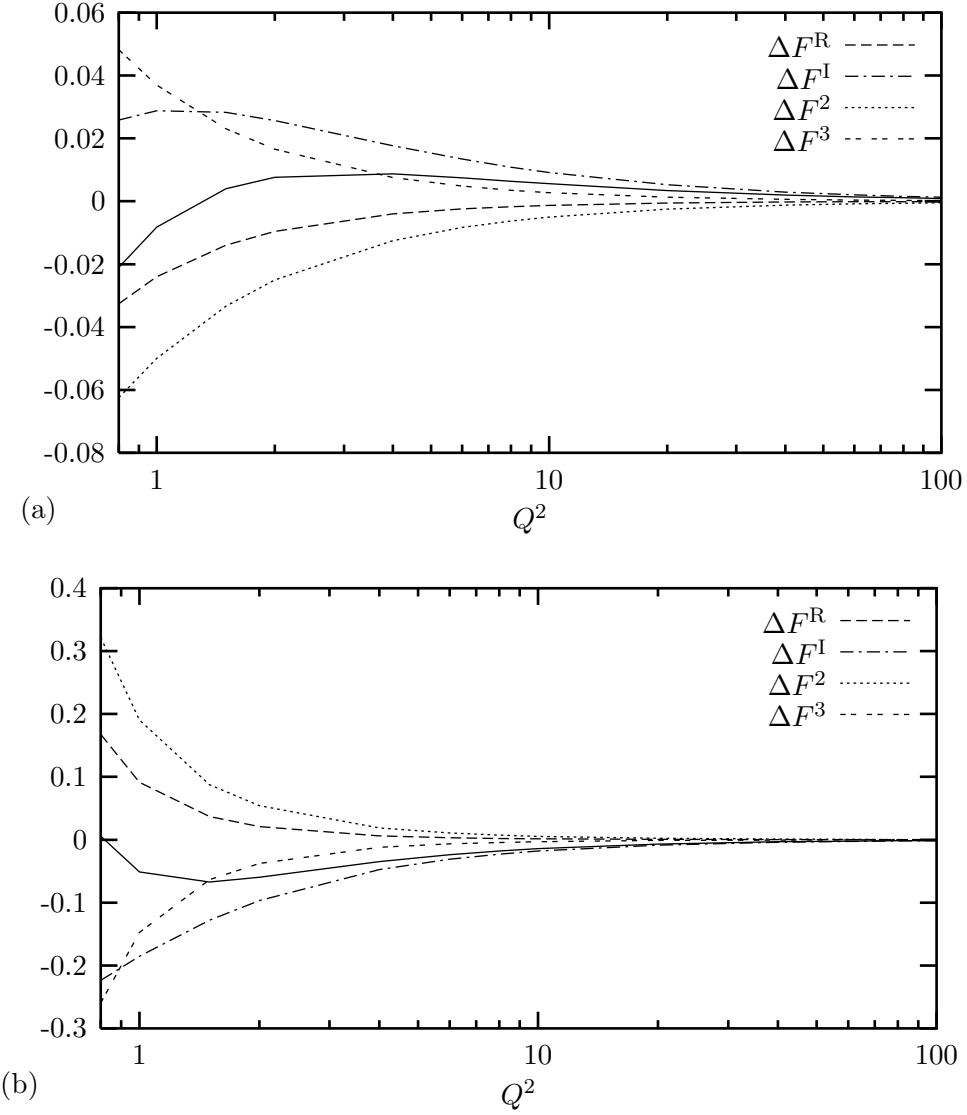}
    \caption{$Q^2$-dependence of the (a) longitudinal and (b) transverse
      twist-four contributions (solid line) and the four terms they are
      composed of at $x=10^{-3}$.
      Normalization with transverse leading-twist values.
      } \label{Const_plot04ab}
  \end{center}
\end{figure}

Finally, figs.\,\ref{Const_plot04ab}a and b show the $Q^2$-dependence of the
longitudinal and transverse twist-four contributions at $x=10^{-3}$. As a
general feature, we observe the interplay between the $1/Q^2$ suppression
factor and the rise in the $Q^2$-evolution. In both figures, the change of the
sign in the sum of all contributions is due to the increase of $\DelFI$.  Since
we have performed our computations in DLA, the various twist-four contributions
either vanish for $Q^2\to Q_0^2$ or become a small constant. Moreover, $\DelFI$
goes much faster to zero than all other contributions. As discussed before,
this is a special problem of the DLA approximation of $\DelFI$.  If we would
take into account also sub-leading terms, this decrease of $\DelFI$ for low
$Q^2$ would be less distinct. It is, therefore, likely that the dominance of
$\DelFI$ in the twist-four corrections to $F_2$ is, in fact, stronger than
found in our DLA analysis.

In summary, in this scenario the transverse twist-four contribution gives a
negative correction of up to fifteen per cent of leading-twist at
$Q^2=1\,$GeV$^2$, while the longitudinal twist-four contribution remains much
smaller.  In both cases we find cancellations between the various twist-four
contributions which enter with different signs. In particular, $\DelFR$
contributes with a rather large amount compared to $\DelFI$. In retrospect it
seems justified that we have taken into account also the twist-four
contributions of $\DelFtwo$ and $\DelFthree$. If we argue that $\DelFthree$
should be related to the next-to-leading order corrections of the BFKL
amplitude, $\DelFR$ would correspond even to NNLO corrections, while $\DelFI$
is the first term which contributes with a stronger power-behavior (cf.\ 
eqn.\,\eqref{D4_Sad_Approx1}).
\begin{figure}
  \begin{center}
    \input{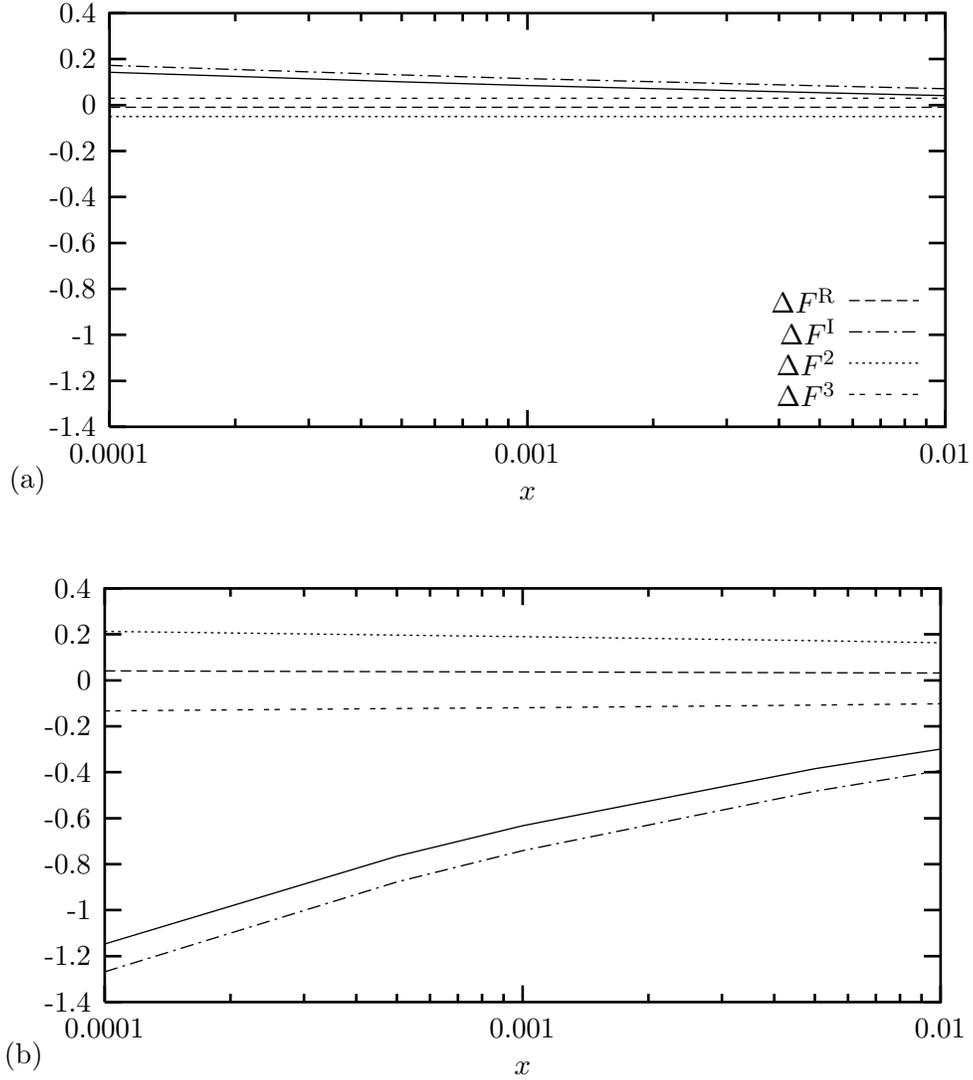}
    \caption{(a) longitudinal and (b) transverse twist-four contributions
      as a function of $x$ at $Q^2=1\,$GeV$^2$. For this scenario, we have
      chosen constant initial distributions with a strong choice of the
      anti-symmetric initial distribution that enters into $\DelFfour$.
      The two diagrams, shown here, correspond to diagrams
      \ref{Const_plot02ab}a and \ref{Const_plot03ab}a in the previous
      scenario. Normalization with transverse leading-twist values.
      } \label{Const_plot_strong01}
  \end{center}
\end{figure}
%
%
\subsubsection{Strong Anti-Symmetric Initial Distribution $(\lambda=1.5)$}
\label{SecStrongAnti}
In this subsection we choose a stronger value for the initial condition
$\phifA$, viz.\ $\phifA=1.5\,\phifS$.  Since $\varphi_4^\mathrm{diff}$ remains
unchanged, our new scenario results in larger values for both $\phifS$ and
$\phifA$, cf.\,\eqref{RelPhiDiff_PhiSA1}.  As we have outlined above, $\DelFI$
couples only to $\varphi_{4\mathrm S}$, while $\DelFR$ couples to both
$\varphi_{4\mathrm S}$ and $\phifA$. From \eqref{deltaFR} we expect that an
increase in $\varphi_{4\mathrm A}$ leads to a larger value of $\DelFI$, while
$\DelFR$ decreases. This is confirmed by the numerical values shown in
fig.\,\ref{Const_plot_strong01} for the longitudinal and transverse
contributions at $Q^2=1\,$GeV$^2$. Both, in the longitudinal and in the
transverse case, $\DelFI$ now dominates. The negative transverse twist-four
correction has now gone up to more than 100\% (at $x=10^{-4}$) of the
leading-twist contribution, whereas the positive longitudinal term reaches
15\%. Certainly, if this scenario is realistic, our consideration of twist-four
is not sufficient. As soon as twist-four reaches, say, fifty per cent of
leading-twist, we have to consider also contributions with twist larger than
four. Using, again, \eqref{deltaFR} we deduce that in this scenario (for the
transverse photon at $Q^2=1\,$GeV$^2$) the contribution due to $\phifA$ reaches
about 130\,\% of leading-twist, i.e.\ $\lambda=1.5$ is likely to present an
upper limit of $\phifA$.

In both scenarios, considered so far, we find sizeable negative twist-four
corrections at $Q^2$ near $1\,$GeV$^2$: this confirms the expectation that the
transition from the parton picture to Regge physics at $Q^2=0$ is indeed
accompanied by sizable higher-twist.  In particular, twist-four is negative at
small $x$.

We believe that the largest uncertainties in this analysis are due to the
unknown initial condition $\phifA$. As we have demonstrated, a rather small
variation in the ratio $\phifA/\phifS$ has a large effect on the relative
contributions of $\DelFI$ and $\DelFR$, and, therefore, on the total twist-four
contributions. It is, therefore, important to find further restrictions on the
size of $\phifA$.  Other sources of possible errors are our choice of the size
of $\DelFthree$, and the double logarithmic approximation.  Moreover, as
outlined above, it seems possible that $\DelFI$ at very low $Q^2$ would even be
larger if one would take into account also sub-leading terms.
\begin{figure}
  \begin{center}
    \input{plot05.pst}
    \caption{(a) longitudinal and (b) transverse twist-four contributions
      as a function of $x$
      at $Q^2=1\,$GeV$^2$. For this scenario, we have
      chosen increasing initial distributions
      ($\omega_0=0.2$) at $Q_0^2=0.8\,$GeV$^2$.
      Normalization with transverse leading-twist values.
      } \label{Incr_plot01}
  \end{center}
\end{figure}
%
%
\subsection{Increasing Initial Parton Distribution}
As an alternative, we have considered a scenario with an initial gluon
distribution which increases at small $x$: $xg(x,Q_0^2)\sim(1/x)^{\omega_0}$.
For the twist-four part of the two-gluon amplitude and for the three-gluon
contribution we assume the same power-behavior of the initial contribution as
in the leading-twist case, and we have set $\omega_0=0.2$. As to the
$\DelFfour$ contributions, the situation is slightly more subtle.  Since the
four-gluon operator in $\DelFI$ is closely related to the double-exchange of
DGLAP ladders, it seems reasonable to assume that the initial distribution of
the four-gluon amplitude behaves like $(1/x)^{2\omega_0}$, i.e.\ the functions
$\varphi_4(\omega)$ in \eqref{initialtotal} are assumed to have a pole at
$2\omega_0$.  For $\DelFR$ there is an ambiguity: on the one hand, the lowest
order diagram is the four-gluon exchange, i.e.\ the evolution should start with
$\varphi_4$ which rises as $\sim(1/x)^{2\omega_0}$. On the other hand, if we
consider $\DelFR$ as a NNLO correction to the two-gluon operator
\eqref{Two_G_Op1}, one might argue that we should have the same initial
conditions as for the twist-four part of the two-gluon amplitude, i.e.\ 
$\sim(1/x)^{\omega_0}$. In our numerical analysis we have chosen the first
option.  Finally, since it seems unrealistic to assume an increasing initial
distribution at $Q_0^2=0.5\,$GeV$^2$, we, now, set $Q_0^2=0.8\,$GeV$^2$.
Moreover we return to our initial convention and set
$\varphi_\mathrm{4S}=\varphi_\mathrm{4A}$.

Fig.\,\ref{Incr_plot01} shows the twist-four contributions at $Q^2=1\,$GeV$^2$
for the longitudinal and transverse photon.  Compared to our first scenario,
the twist-four corrections are, now, much larger. In order to understand this
growth of the twist-four corrections we compare the first scenario presented in
figs.\ref{Const_plot02ab}a and \ref{Const_plot03ab}a. Due to the increasing
initial distribution, all four contributions are now larger than in the case of
constant initial distributions.  In particular, $\DelFtwo$ and $\DelFthree$ are
magnified by a similar factor, and the approximate cancellation works in the
same way as before. For $\DelFR$ and $\DelFI$, on the other hand, the
approximate balance is distorted: at $Q^2=1\,$GeV$^2$ we are close to the
initial scale, and $\DelFI$ is still small. Moreover, because of our choice of
the $1/x$-power $\DelFR$ has a stronger rise in $1/x$ than $\DelFtwo$ and
$\DelFthree$ and, therefore, dominates. As result, the twist-four corrections
at $Q^2=1\,$GeV$^2$ are dominated by $\DelFR$ which is, now, much less balanced
than in the previous scenario. By choosing in $\DelFR$ a weaker $x$-dependence
at the initial scale and by setting our initial scale, as before, to
$Q_0^2=0.5\,$GeV$^2$, it may be possible to partly restore the balance and to
obtain a somewhat smaller twist-four correction. This scenario, therefore,
represents a somewhat extreme possibility.

Nevertheless, the trend seen in this third scenario may be characteristic for
increasing initial distributions.  Initial parton distributions which increase
at small $x$ tend to increase the twist-four corrections at low $Q^2$, in
particular through $\DelFR$.  This pushes the balance between the positive
$\DelFR$ and the negative $\DelFI$ into the positive direction, and, as in our
case, may even lead to a net positive twist-four correction to the
leading-twist structure function.  If these tentative conclusions are correct,
we have to interpret the scenario with rising initial conditions as being
unrealistic.  As discussed in the beginning of this paper, twist-four
corrections could, in principle, provide an explanation for the observed change
in the small $x$ behavior at low $Q^2$.  In particular, with decreasing $Q^2$,
a negative transverse higher-twist could, because of its stronger rise at small
$x$, overcome the $1/Q^2$ suppression and interfere with the rising
leading-twist contribution. In this last scenario, however, we found that the
opposite sign for transverse higher-twist is more likely, and leading- and
higher-twist are adding up rather than compensating each other at low $Q^2$. We
believe that such a scenario is not very likely to be realistic.
%
%
\section{Conclusions}
In this paper we have made a first attempt to estimate twist-four corrections
to the gluon structure function at small $x$ and low $Q^2$. We have collected
all presently available analytic information on the twist-four gluon evolution
equations and defined the terms which need to be included. Up to now DLA is the
only approximation for which explicit expressions exist.  One of the main
features of gluonic twist-four corrections is the sign structure: we find, both
in the transverse and in the longitudinal structure functions, terms with
opposite signs which tend to compensate each other.

In the second part we have investigated how DIS diffractive dissociation (in
particular the production of longitudinal vector mesons) can be used for
constraining the free parameters, i.e.\ the initial conditions. As a main
result, the AGK rules provide a bridge between DIS diffractive dissociation and
the twist-four part of the four-gluon amplitude, but there remains a piece in
the initial condition which cannot be determined in this way.

In our numerical part we have considered three different scenarios.  For two of
them we have chosen constant (in $x$) initial parton distributions at the input
scale $Q_0^2$, and we have varied the relative strength of pieces in the input
distribution. In addition, we have studied also one scenario in which the input
distribution rises in $1/x$.  In general we find, both in the transverse and in
the longitudinal structure functions, significant cancellations between terms
with opposite signs which tend to weaken or even compensate each other.
Moreover, the transverse twist-four corrections are larger than the
longitudinal ones. For the scenarios with constant initial conditions the sum
of the transverse twist-four corrections is negative, and at small $x$ grows
stronger than leading-twist. For $Q^2=1\,$GeV$^2$, $x=10^{-4}$ twist-four
ranges between 15\% and 130\% of leading-twist. Compared to the results of
\cite{Caldwell}, our corrections seem to become significant in the right
region.  For the case of increasing initial conditions the different pieces of
the transverse twist-four contributions add up to a positive higher-twist
correction; we interpret this scenario as rather unrealistic. This may however
change if one could go beyond the DLA\@. All our numerical results are based
upon DLA accuracy and, therefore, should be taken as only ``semiquantitative''.

Before firm conclusions can be drawn, several improvements have to be made.
First of all, we need the complete LO calculations of the evolution kernels of
the gluonic operators.  Work along these lines is in progress \cite{BBS}, but
it will take some time before one can start to perform numerical calculations.
Secondly, as can be seen already from eqn.\,\eqref{D4final}, the amount of
computer work necessary to calculate the $Q^2$ evolution of the gluon operators
is substantially larger than in the leading-twist case. Presently, it is not
clear whether it is possible to find a closed expression also beyond DLA or
whether we have to solve evolution equations in $x$-space.  Attempts to solve
the set of coupled evolution equations \eqref{Bet_Salp1} will face difficulties
with storage capacities. Finally, as discussed above, experimental information
on specific final states is needed in order to restrict the arbitrariness in
the choice of the initial conditions. In this first attempt we have used only
the diffractive longitudinal vector meson production in order to obtain a first
hint on the magnitude of twist-four.  Besides this, other information of final
states might be useful; a promising candidate might be a measurement of the
two-particle correlation function as a function of rapidity \cite{CL}.

{\bf Acknowledgements:} One of us (J.B.) gratefully acknowledges the
hospitality of the HEP Theory Division of Argonne National Laboratory and of
the Theory Division of Fermilab where part of this work has been done.
Discussions with R.K.\,Ellis have been very helpful and encouraging.
%
%
\begin{appendix}
\section*{Appendix}
\setcounter{section}{1}
We, first, give expressions for the vectors in color space $\CLket{i}_\chi$ in
the coupling scheme $\chi=(1234)$:
\begin{eqnarray}
  \CLket{1}_{1234} &=& \frac18 \delta^{ab}\delta^{cd} \\
  \CLket{8_\mathrm{A}}_{1234} &=& \frac{-1}{6\sqrt2} f^{abl}f^{lcd} \\
  \CLket{8_\mathrm{S}}_{1234} &=& \frac3{10\sqrt2} d^{abl}d^{lcd} \\
  \CLket{10+\bar{10}}_{1234} &=& \frac1{2\sqrt5} \left\{
    \frac12[\delta^{ad}\delta^{bc}-\delta^{ac}\delta^{bd}] +
    \frac13 f^{abl}f^{lcd} \right\} \\
  \CLket{27}_{1234} &=& \frac1{3\sqrt3}  \left\{
    \frac12[\delta^{ad}\delta^{bc}+\delta^{ac}\delta^{bd}] -
    \frac18\delta^{ab}\delta^{cd}
    -\frac35 d^{abl}d^{lcd} \right\} \, .
\end{eqnarray}
They are normalized to one
\begin{equation}
  _\chi\!\CLbraket{i}{j}_\chi = \delta^{ij}
\end{equation}
and eigenstates of the BFKL color factor $t_{2\to2}$:
\begin{equation}
  (t_{2\to2})_{\alpha\alpha'}\CLket{i}_{\alpha\alpha'\beta\beta'}=
    \varepsilon_i \CLket{i}_{\alpha\alpha'\beta\beta'} \qquad
  (t_{2\to2})_{\beta\beta'}\CLket{i}_{\alpha\alpha'\beta\beta'}=
    \varepsilon_i \CLket{i}_{\alpha\alpha'\beta\beta'} \, ,
\end{equation}
where
\begin{equation}
  (t_{2\to2})_{12}=f^{aa'l}f^{lb'b} \qquad (t_{2\to2})_{13}=f^{aa'l}f^{lc'c}
  \qquad \mathrm{etc.}
\end{equation}
and $\varepsilon_i=\{-3,-\frac32,-\frac32,0,1\}$ for $i=\{1,8_\mathrm
A,8_\mathrm S,10+\bar{10},27\}$. They are related to vectors in other coupling
schemes via transitions with the matrices
\begin{equation}
  \Lambda=\begin{pmatrix}
    \frac18 & -\frac1{2\sqrt2} & \frac{1}{2\sqrt{2}} & 
      -\frac{\sqrt5}4 & \frac{3\sqrt{3}}{8} \\
    -\frac1{2\sqrt2} & \frac12 & -\frac12 & 0 & \frac12\sqrt{\frac32} \\
    \frac{1}{2\sqrt{2}} & -\frac12 & -\frac{3}{10} & 
      \sqrt{\frac25} & \frac{3}{10}\sqrt{\frac32} \\
    -\frac{\sqrt5}4 & 0 & \sqrt{\frac25} & \frac12 & \frac14\sqrt{\frac35} \\
    \frac{3\sqrt{3}}{8} & \frac12\sqrt{\frac32} & \frac{3}{10}\sqrt{\frac32} & 
      \frac14\sqrt{\frac35} & \frac{7}{40}
  \end{pmatrix}
  \qquad \mathrm{and} \qquad
  P=\mathrm{diag}(1,-1,1,-1,1)
\end{equation}
in the following way:
\begin{align}
  \CLket{i}_{1234} &= \sum_j \Lambda_{ij} \CLket{j}_{1324} &
  \CLket{i}_{1324} &= \sum_j \Lambda_{ij} \CLket{j}_{1234} \notag\\
  \CLket{i}_{1234} &= \sum_j (P\Lambda P)_{ij} \CLket{j}_{1423} &
  \CLket{i}_{1423} &= \sum_j (P\Lambda P)_{ij} \CLket{j}_{1234} \\
  \CLket{i}_{1324} &= \sum_j (P\Lambda)_{ij} \CLket{j}_{1423} &
  \CLket{i}_{1423} &= \sum_j (\Lambda P)_{ij} \CLket{j}_{1324} \, .\notag
\end{align}

Next, we give explicit expressions for the components of vector
\begin{equation}
  \bm{\chi}(\omega\tilde\nu) \equiv \Sigma(\omega\tilde\nu) \, \bm{V}
\end{equation}
in eqns.\,\eqref{D4_eqn1} and \eqref{D4final}. The results are the same for the
three coupling schemes. Therefore, we list the components of $\bm{\chi}$ for
the three contributing color states and write them as functionals of the
auxiliary quantities $f_i(\omega\tilde\nu)$, $i=1\dots 6$, and
$g_j(\omega\tilde\nu)$, $j=1\dots 7$.  They are defined as follows:
\begin{align}
  f_1 &=-\frac{1}{32}\gamma_1 + \frac3{40}\gamma_8-
      \frac{27}{160}\gamma_{27}-\frac14
    & f_4 &=\frac1{\gamma_1}-\frac1{128}\gamma_1-\frac1{16}\gamma_8-
      \frac{27}{128}\gamma_{27}-\frac1{16}\notag\\
  f_2 &=-\frac1{128}\gamma_1-\frac1{80}\gamma_8-
      \frac7{640}\gamma_{27}-\frac1{16}
    & \qquad 
    f_5 &=\frac1{\gamma_8}-\frac1{16}\gamma_1-\frac9{200}\gamma_8-
      \frac{27}{400}\gamma_{27}+\frac3{20}\\
  f_3 &=-\frac3{32}\gamma_1+\frac9{200}\gamma_8-
       \frac{21}{800}\gamma_{27}-\frac3{20}
    & f_6 &=\frac1{\gamma_{27}}-\frac{27}{128}\gamma_1-\frac{27}{400}\gamma_8-
      \frac{49}{3200}\gamma_{27}-\frac7{80} \notag
\end{align}
and
\begin{align}
    g_1 &=\frac{\sqrt3}2 f_1\cdot f_3 - 3\sqrt3 f_2\cdot f_5
      \qquad\qquad\qquad\: g_4 =f_5\cdot f_6 - \frac32 f_3^2 \notag\\
    g_2 &= \frac{9\sqrt2}2 f_2\cdot f_3 -\frac{\sqrt2}2 f_1\cdot f_6
      \qquad\qquad\qquad  g_5 =f_4\cdot f_6 - 27 f_2^2 \\
    g_3 &=\frac{3\sqrt6}2 f_1\cdot f_2 - \frac{\sqrt6}2 f_3\cdot f_4
      \qquad\qquad\qquad  g_6 = f_4\cdot f_5 - \frac12 f_1^2 \notag\\
    g_7 &= 9f_1\cdot f_2\cdot f_3 + f_4\cdot f_5\cdot f_6 -
           \frac12 f_1^2\cdot f_6 -
      27f_2^2\cdot f_5 - \frac32 f_3^2\cdot f_4 \, , \label{g7_def1}
\end{align}
where the functions $\gamma_i(\omega\tilde\nu)$ are the diagonal elements of
matrix $\gamma(\omega\tilde\nu)$ in eqn.\,\eqref{gam_and_lam_def1}, and we have
used $\gamma_8$ instead of $\gamma_{8_\mathrm{S}}$. With these functions the
elements of the vector $\bm{\chi}$ are given by the following expression:
\begin{equation} \label{VecCompRes1}
  g_7\cdot \chi_j = v_j g_7\gamma_j +
  \sum_{i=1}^6 g_i\left( \kappa_{1i}^{(j)} \gamma_j\gamma_1 +
    \kappa_{2i}^{(j)} \gamma_j\gamma_8 + \kappa_{3i}^{(j)} \gamma_j\gamma_{27}+
    \kappa_{4i}^{(j)} \gamma_1 + \kappa_{5i}^{(j)} \gamma_8 +
    \kappa_{6i}^{(j)} \gamma_{27} \right) \, ,
\end{equation}
where $j\in \{1,8_\mathrm{S},27\}$, $v_1=2$, $v_8=4\sqrt2$, $v_{27}=6\sqrt3$
(cf.\,\eqref{Vcomp1}) and the coefficients $\kappa_{li}^{(j)}$ are listed in
table~\ref{Exp_Tab1}. The pole at $\omega\tilde\nu=4(1+\delta)N_\mathrm
c\alphaS/\pi$ (eqn.\,\eqref{Pol_Pos1}) is due to a zero of function $g_7$,
eqn.\,\eqref{g7_def1}.

Finally, the color tensor $d^{abcd}$ in \eqref{D4R} has the form \cite{BW}:
\begin{eqnarray}
  d^{abcd} &=& \CLtr\{t^{a}t^{b}t^{c}t^{d}\}+
    \CLtr\{t^{d}t^{c}t^{b}t^{a}\} =
  \nonumber \\ &=&
  \frac{1}{6}\delta^{ab}\delta^{cd}
    + \frac{1}{4} d^{abl}d^{lcd}
    - \frac{1}{4} f^{abl}f^{lcd}
  \nonumber \\ &=&
  \frac{1}{6}\delta^{ad}\delta^{bc}
    + \frac{1}{4} d^{adl}d^{lbc}
    + \frac{1}{4} f^{adl}f^{lbc} \,,
\end{eqnarray}
where the Gell-Mann matrices $t^{a}$ are normalized to 
\begin{equation}
  \CLtr\{t^{a}t^{b}\}=\frac{1}{2} \delta^{ab} \,.
\end{equation}
\end{appendix}
%
%

%
%
\begin{table}
  \begin{center}
    \begin{tabular}{c|rrrrrr}
      $i$ & 1 & 2 & 3 & 4 & 5 & 6 \\[1pt]
      \hline
      \\[-7pt]
      $\kappa_{1i}^{(1)}$ & $\frac{3\sqrt3}{32}$ & $\frac{\sqrt2}{16}$ &
        $\frac{3\sqrt6}{16}$ & $\frac1{64}$ & $\frac18$ & $\frac{27}{64}$ 
        \\[5pt]
      $\kappa_{2i}^{(1)}$ & $\frac{9\sqrt3}{20}$ & $\frac{7\sqrt2}{40}$ &
        $-\frac{3\sqrt6}{40}$ & $\frac18$ & $-\frac3{10}$ & $\frac{27}{40}$
        \\[5pt]
      $\kappa_{3i}^{(1)}$ & $\frac{213\sqrt3}{160}$ & $\frac{81\sqrt2}{80}$ &
        $\frac{51\sqrt6}{80}$ & $\frac{27}{64}$ & $\frac{27}{40}$ & 
        $\frac{189}{320}$ \\[5pt]
      $\kappa_{4i}^{(1)}$ & $\frac{3\sqrt3}4$ & $\frac{\sqrt2}2$ && $\frac14$
      \\[5pt]
      $\kappa_{5i}^{(1)}$ & $\frac{6\sqrt3}5$ & $-\frac{6\sqrt2}5$ && $2$
      \\[5pt]
      $\kappa_{6i}^{(1)}$ & $\frac{21\sqrt3}{20}$ & $\frac{27\sqrt2}{10}$
        && $\frac{27}4$ \\[5pt]
      \hline
      \\[-7pt]
      $\kappa_{1i}^{(8)}$ & $\frac{9\sqrt6}{80}$ & $\frac7{80}$ &
        $-\frac{3\sqrt3}{80}$ &
        $\frac{\sqrt2}{32}$ & $-\frac{3\sqrt2}{40}$ & $\frac{27\sqrt2}{160}$
        \\[5pt]
      $\kappa_{2i}^{(8)}$ & $\frac{3\sqrt6}{10}$ & $-\frac35$ &
        $-\frac{9\sqrt3}{25}$ &
        $\frac{\sqrt2}4$ & $\frac{9\sqrt2}{50}$ & $\frac{27\sqrt2}{100}$
        \\[5pt]
      $\kappa_{3i}^{(8)}$ & $\frac{51\sqrt6}{80}$ & $-\frac{27}{80}$ &
        $\frac{99\sqrt3}{400}$ & $\frac{27\sqrt2}{32}$ &
        $-\frac{81\sqrt2}{200}$ & $\frac{189\sqrt2}{800}$ \\[5pt]
      $\kappa_{4i}^{(8)}$ && $\frac14$ & $\frac{3\sqrt3}4$ && $\frac{\sqrt2}2$
      \\[5pt]
      $\kappa_{5i}^{(8)}$ && $2$ & $\frac{6\sqrt3}5$ && $-\frac{6\sqrt2}5$
      \\[5pt]
      $\kappa_{6i}^{(8)}$ && $\frac{27}4$ & $\frac{21\sqrt3}{20}$ &&
      $\frac{27\sqrt2}{10}$ \\[5pt]
      \hline
      \\[-7pt]
      $\kappa_{1i}^{(27)}$ & $\frac{71}{160}$ & $\frac{9\sqrt6}{80}$ &
      $\frac{17\sqrt2}{80}$ &
        $\frac{3\sqrt3}{64}$ & $\frac{3\sqrt3}{40}$ & $\frac{21\sqrt3}{320}$
        \\[5pt]
      $\kappa_{2i}^{(27)}$ & $\frac{17}{20}$ & $-\frac{3\sqrt6}{40}$ &
      $\frac{33\sqrt2}{200}$ &
        $\frac{3\sqrt3}8$ & $-\frac{9\sqrt3}{50}$ & $\frac{21\sqrt3}{200}$
        \\[5pt]
      $\kappa_{3i}^{(27)}$ & $\frac{189}{160}$ & $\frac{81\sqrt6}{80}$ &
      $\frac{189\sqrt2}{400}$ &
        $\frac{81\sqrt3}{64}$ & $\frac{81\sqrt3}{200}$ &
        $\frac{147\sqrt3}{1600}$ \\[5pt]
      $\kappa_{4i}^{(27)}$ & $\frac14$ && $\frac{\sqrt2}2$ &&&
      $\frac{3\sqrt3}4$ \\[5pt]
      $\kappa_{5i}^{(27)}$ & $2$ && $-\frac{6\sqrt2}5$ &&& $\frac{6\sqrt3}5$
      \\[5pt]
      $\kappa_{6i}^{(27)}$ & $\frac{27}4$ && $\frac{27\sqrt2}{10}$ &&&
      $\frac{21\sqrt3}{20}$ \\[5pt]
    \end{tabular}
    \caption{Expansion coefficients for the three different 
      color states in eqn.\,\eqref{VecCompRes1}.}
    \label{Exp_Tab1}
  \end{center}
\end{table}
\end{document}